\documentclass[iop,apj]{emulateapj}

\usepackage{amsmath}
\usepackage{mathrsfs}
\usepackage{color}

\usepackage{times}
\usepackage{color}

\usepackage{tabularx}
\usepackage[multidot]{grffile}
\usepackage[roman]{parnotes}
\usepackage[table]{xcolor}
\usepackage{xcolor}

\include{boldgreek}

\shorttitle{Elementary Events in the Solar Corona}
\shortauthors{Einaudi {\it et al.}}

\begin{document}

\title{Energetics and 3-D Structure of Elementary Events in Solar Coronal Heating}

\author{G. Einaudi$^1$, R. B. Dahlburg$^{2}$, I. Ugarte-Urra$^3$, J. W. Reep$^3$, A. F. Rappazzo$^4$, and M. Velli$^4$}
\affil{\vspace{.2em}
$^1$Department of Physics and Astronomy, George Mason University, Fairfax, VA 22030, USA\\
$^2$LCP\&FD, Naval Research Laboratory, Washington, DC 20375, USA; rdahlbur@lcp.nrl.navy.mil \\
$^3$Space Science Division, Naval Research Laboratory, Washington, DC 20375, USA\\
$^4$EPSS, UCLA, Los Angeles, CA 90095, USA}

\begin{abstract}
\cite{1972ApJ...174..499P}  first proposed that coronal heating was the necessary outcome of an energy flux caused by the tangling of coronal magnetic field lines by photospheric flows. In this paper we discuss how this model has been modified by subsequent numerical simulations outlining in particular the substantial differences between the ``nanoflares" introduced by Parker and  ``elementary events", defined here as small-scale spatially and temporally isolated heating events resulting from the continuous formation and dissipation of field-aligned current sheets within a coronal loop. We present numerical simulations of the compressible 3-D MHD equations using the HYPERION code.  We use two clustering algorithms to investigate the properties of the simulated elementary events: an IDL implementation of a Density-Based Spatial Clustering of Applications with Noise (DBSCAN) technique; and our own Physical Distance Clustering (PDC) algorithm. We identify and track elementary heating events in time, both in temperature and in Joule heating space. For every event we characterize properties such as: density, temperature, volume, aspect ratio, length, thickness, duration and energy.  The energies of the events are in the range $10^{18}-10^{21}$ ergs, with durations shorter than 100 seconds. A few events last up to 200 seconds and release energies up to $10^{23}$ ergs. While high temperature are typically located at the flux tube apex, the currents extend all the way to the footpoints. Hence a single elementary event cannot at present be detected. The observed emission is due to the superposition of many elementary events distributed randomly in space and time within the loop.
\end{abstract}


\section{Introduction}
Loops in the solar corona are characterized by the steady and diffuse emission of X-rays and Extreme-Ultraviolet radiation.
The coronal temperatures implied by this radiation indicate that in some way the corona is heated to temperatures significantly greater than
the underlying chromosphere.  Many different scenarios are at present under investigation to explain this heating.
In all cases the ultimate energy source derives from convective motions in which there is more than enough energy to supply total coronal losses, the main question being how the energy is deposited in corona. In the last few decades it has become more and more evident that the heating process involves the continuous formation and disruption of many small scale current sheets.  In each current sheet the energy is dissipated and transformed into internal energy with a consequent formation of temperature peaks. This local phenomenon can be termed the "elementary event" underlying coronal heating \citep{Chiuderi 1993}.  In this paper we present numerical evidence relating to the extent, duration and energetics of these elementary events.

In this introductory section we detail the history of the "elementary event" concept beginning with the work of
 \cite{1972ApJ...174..499P}, who was the
first to explore in detail the \cite{Gold & Hoyle 1960} idea that coronal heating could be the necessary outcome of an
energy flux associated with the tangling of coronal field
lines by photospheric flows. \cite{Parker 1983}, \cite{1988ApJ...330..474P}, \cite{Sturrock &
Uchida 1981}, \cite{Van Ballegooijen 1986}, \cite{GFF} and \cite{Berger 1991},
among others, further explored the dynamics caused by
such random shuffling of magnetic field lines, using either weakly nonlinear analyses, turbulence phenomenologies and
or extended dimensional arguments.
The three-dimensional simulations performed by \cite{Mikic 1989},
\cite{Strauss 1993}, \cite{Longcope & Sudan 1994}, \cite{1996JGR...10113445G}, and
\cite{Hendrix & Van Hoven 1996} confirmed
that this process causes the formation and exponential growth of local coronal currents.

To perform moderate- to high-resolution simulations of the system, allowing for much larger integration
times, \cite{1996ApJ...457L.113E} developed a
two-dimensional model, focusing on the dynamics in a given plane.
The photospheric forcing function used to feed energy into the
plasma has large-scale spatial structures (with amplitudes changing at
the photospheric characteristic convection timescale) so that
small-scales are not directly excited.
Therefore, in the 2-D simulations, a magnetic forcing is imposed varying on the typical photospheric time scale $\tau_p$,
a first attempt at
studying  the Parker scenario with significant resolution, in the effort to detail
the perpendicular dynamics without any assumptions on the
properties of the resulting current sheets \citep{1996ApJ...457L.113E,Georgoulis,Dmitruk 1998}.
The transverse
dynamics for times longer than $\tau_p$ was shown to alter  the magnetic field structure
substantially from the spatial structure of the magnetic forcing, and
the system evolution is turbulent and highly dynamical, exhibiting intermittency for
both the spatial current distribution and the mean dissipation time series.
Such simulations strongly  suggest  that
the corona is heated by the dissipation processes occurring in a large number of small scale current sheets.
the  corona  is built  up by a large number  of small scale current sheets  where dissipative processes take  place.

Based on these results, a series of 3-D numerical simulations solving the simplified reduced magnetohdrodynamic (RMHD) equations, introduced  by \cite{Strauss 1976}, in Cartesian geometry were performed.
The goal of these simulations was twofold: 1) to determine how a coronal loop responds to different photospheric velocity patterns, and
2) to investigate how the electric current sheets are formed as well as the details of the heating which occurs through the dissipation of magnetic energy
\citep{2007ApJ...657L..47R, 2008ApJ...677.1348R, 2010ApJ...722...65R, 2011PhRvE..83f5401R, 2013ApJ...773L...2R, 2015ApJ...815....8R}.
The results detailed in these papers confirmed the 2-D results that the energy flux entering the corona because of photospheric motions causes a turbulent cascade of energy towards small scales, with electric current sheets continuously being created and destroyed throughout the coronal loop.
The turbulence that develops is highly intermittent.  The dissipation of energy occurs in current sheets, which are localized structures.
Hence the heating also occurs on small spatial scales (Note that there is also particle acceleration, although that is beyond the scope of the present model.).  When the loop is fully turbulent it achieves a statistically steady state in which, on average, the Poynting flux induced at the the boundaries by footpoint convection is balanced by the dissipation of energy in the electric current sheets.  In this steady state, saturation occurs for the fluctuating magnetic energy, kinetic energy, mean square electric current, enstrophy and other quantities which are then seen to fluctuate around their mean values.
When the system becomes nonlinear, its behavior is highly dynamical and chaotic.
This state occurs independently of the detailed form of the boundary velocity.  A statistically steady state is achieved in which, although the magnetic field footpoints are convected
continuously by the boundary flows, the resulting topology of the total magnetic field is not simply
a mapping of those flows.  By the same token, the turbulent dynamics that occur are due to
the inherent nonlinear nature of the system, rather than being a consequence of the complexity of the magnetic field
foot point configuration.

For instance a sheared boundary forcing constant in time \citep{2010ApJ...722...65R} or a boundary velocity made of distorted vortices constant in time \citep{2008ApJ...677.1348R} generate nonlinear dynamics similar to that of distorted vortices changing in time on timescales larger than the coronal loop Alfv\'en crossing time \citep[typical of X-ray bright solar loops,][]{2018MNRAS...submitted}. However, static large-scale boundary forcing does not change directly the mapping of the orthogonal magnetic field lines nor create directly small scale structures (and thus they do not force reconnection to occur), rather once the orthogonal magnetic field intensity grows beyond a certain threshold nonlinear dynamics develop because the coronal field is increasingly out of equilibrium in the vortices case \citep{1994ISAA....1.....P, 2013ApJ...773L...2R, 2015ApJ...815....8R}, or because an instability is triggered in the shear case (after which nonlinear dynamics is similar to the other cases with the previous instability not occurring again).

In the fully developed stage, energy which is injected into the system at low spatial wavenumbers
is redistributed by nonlinear interactions through MHD turbulent direct and inverse cascades.
An inertial range develops in both the kinetic energy and magnetic energy.
The spectra of these inertial ranges exhibit a power-law behavior.
Note as well that the magnetic energy tends to be much larger than the kinetic energy.
In configuration space the redistribution of energy has morphological consequences for both lower and higher spatial wavenumbers.
The magnetic field comes to be organized in large scale magnetic islands.
These islands are separated by small scale electric current sheets which extend along the axial magnetic field.
Below a computationally very demanding threshold, but still small respect to the coronal Reynolds numbers
(viz., a magnetic Reynolds of about 800, corresponding to a linear grid resolution of about 8\,km; in this manuscript we reach a resolution of about 15\,km),
average dissipation increases with increasing Reynolds number, as well as peak to average energy dissipation. Beyond this threshold average dissipation is independent of the Reynolds numbers \citep{2007ApJ...657L..47R, 2008ApJ...677.1348R}. The key point is that the energy input to the system is not solely due to the forcing at the boundary, {\it i.e.,}the photospheric velocity, but also to the instantaneous state of the system, {\it i.e.,} the configuration of the coronal magnetic field.  Since the energy input is dependent on both the external forcing and the internal dynamics, the corona is a self-regulating forced system.

The recognition that the Reynolds number is very large in the corona, and the consequent continuous formation and disruption of current sheets where energy is dissipated, leads to a scenario which is qualitatively
similar to the one proposed  by Parker, {\it viz.,} coronal heating is the observable signature of swarms of unobserved individual elementary events.
Parker used order-of-magnitude arguments to estimate the magnitude and the time scale of production of the
mean transverse component of the coronal magnetic field.  In doing so he neglected the internal dynamics of the system which
are driven by the forcing at the boundaries.  From a quantitative point of view, by spatially averaging over the entire system as well as temporally averaging over a time that is long compared with the dynamical time, he derived the amount of energy liberated by magnetic dissipation in many ``elementary events."
The energy release that he computed was about $10^{17}$ Joules.  He termed this averaged energy release event the ``nanoflare''. It is evident that the energy content of Parker ``nanoflares'' is an extreme upper limit of the energy content of an ``elementary event''.

While 2D MHD and RMHD do an excellent job of elucidating the dynamics of coronal heating, they cannot provide any
information of the thermodynamical aspects.The energy due to Ohmic and viscous dissipation is lost from the system since there is no
accounting for thermal energy.  Any effects due to thermal conduction and optically thin radiation are absent, and any feedback
from the thermodynamics on the dynamics is neglected.  The situation is reversed in one-dimensional hydrodynamic models (\cite{2014LRSP...11....4R}.
These models are often used to study the equilibrium state of heated loops, or the flows due to time-dependent evaporation or condensation flows.
What is lacking in these one-dimensional models is the important role of perpendicular dynamics.  Hence an {\it ad hoc} heating function is used to mimic the thermal energy deposition.

Simulation of the coronal heating problem requires a model that can provide a complete reproduction of the coronal heating energy cycle.  Energy is injected in the corona through
the convection of the magnetic field line foot points.  When the perpendicular magnetic field in the corona attains finite amplitude intermittent turbulence develops in which nonlinear interactions lead to direct and inverse cascades.  Energy dissipation is localized in electric current sheets in  which magnetic energy is transformed into thermal energy (and fluctuating magnetic and kinetic energy as well).  Thermal conduction efficiently transfers heat to the chromosphere where it is lost by means of optically thin radiation, and may lead to chromospheric evaporation.
To obtain this model of the full energy cycle it is therefore crucial to augment the previously discussed numerical simulation models with an energy equation that accounts for thermal conduction and optically thin radiation.
In addition, rather than an {\it ad hoc} ``heating'' term, the energy equation should include specific Joule heating and viscous heating terms.
When gravitation effects are considered, it is also necessary to augment the momentum equation with pressure gradient and gravity terms as well as adding in a density equation to account for the effects of stratification.
This augmented system is implemented numerically in the HYPERION code, which solves the full set of compressible magnetohydrodynamic equations \citep{Dahlburg 2012, 2015ApJ...submitted, 2018ApJ...868.116D} (See Section 3).
HYPERION allows for the carrying out of simulations where foot point shuffling heating causes heating due solely to the resistive and viscous dissipation induced in the corona.
With the help of an atomic database (e.g., CHIANTI), temperatures and densities can then be used to synthesize observables, such as the emission of spectral lines, and construct differential emission measures, both useful to examine the radiative consequences of coronal heating.

Note that this added complexity comes with a cost - more variables and more equations imply that this system of equations is computationally more demanding. Hence the range of Lundquist numbers which is achievable is somewhat less than for the 2D MHD and RMHD cases.

HYPERION is typically run with a chromospheric base of dimensions $4000\times4000$ km$^2$. The numerical simulations of our Cartesian coronal loop model neglects the geometry and topological complexity of the solar magnetic field, but allows to use higher numerical resolutions respect to models that take into account entire active regions ({\it e.g.,} \cite{2002ApJ...572L.113G, Bingert, HanHa10, rem17, baumann2013, HanGu15, GuHan17} and references therein), thus enabling a more detailed study of the energy dissipation processes at small scales and its thermodynamical and observational consequences that we describe in the following sections.
The dynamics seen in the RMHD simulations is reproduced.
At any given time a multitude of dissipative current sheets, which are elongated along the axial magnetic field, is observed to form
\citep[a visualization of these current sheets is found in][]{2008ApJ...677.1348R}.
As for the temperature, the loop is found to be a multi-thermal system in which there is a spatial correlation between the electric current sheets and the temperature peaks.
In general, high temperature regions correspond to the locations of the electric current sheets.  An important qualification is that the temperature,
unlike the electric current, also depends on the density.  Due to stratification, the loop is more dense at its footpoints than at its apex.  Hence, while the heating might be the same at both these locations, the temperature near the footpoints does not increase as much. Thus we see that, while the electric current sheets extend from one footpoint to the other, the high temperature regions are more localized with respect to the loop apex.
If the parameters of the loop, such as it's length, magnetic field strength,
or the boundary velocity are varied, then important features of the temperature
distribution, such as the peak, extent, and duration, are found to vary.

The turbulent character of the heating has important implications for observations.
The distribution of heating temperature in the loop is inhomogeneous,
Furthermore both spatial and temporal intermittency must be taken into account to interpret observations
correctly.
hence significant heating occurs in only a small part of the loop volume at any given time.
Thus high-temperature plasma radiation originates from a very small fraction of the
loop volume with a mass much lower than the total mass of the loop.
These temperature structures cannot, at present, be resolved observationally due to their
small spatial extent.
Using the computed number densities and temperatures, \cite{2015ApJ...submitted} performed an
emission measure analysis of their simulated loops.
A loop length of 50,000 km and axial magnetic fields of 0.01,
0.02, and 0.04 Tesla were examined.
The computational emission measures derived from the simulated intensities were found to be in excellent
agreement with emission line intensities observed by the Extreme Ultraviolet (EUV) Imaging
Spectrometer (EIS, \citealt{2007SoPh..243...19C}) would observe.
Furthermore, in spite of the strong spatial and temporal intermittency of the loop system, the computed emission measures
were found to retain the same form for the duration of the numerical simulations.

\cite{2018ApJ...868.116D} studied the observational consequences for the temperature as the loop magnetic field strength is increased and/or the length of the loop decreases.
This is equivalent to decreasing the Alfv\'en crossing ($\tau_A$) time of the loop.
It was found that as $\tau_A$ decreases, both the peak temperature and the width of the temperature distribution of the loop exhibit increases.
The emission measure broadening implies that the loop is more and more becoming a multi-temperature system.
Let $L_0$ denote the length of the coronal loop and $B_0$ the magnetic field magnitude of the loop.
The predictions of how the effective temperature dependence varies with respect to the $B_0/L_0$ ratio are in agreement with the observational estimates of those quantities obtained by \citet{2017ApJ...842...38X}, who examined fifty coronal loop in the $<$1--2 MK range.
A larger observational temperature range is needed to determine the slope of the temperature dependence.
Proportionality between temperature and emission width has been observed in coronal loops \citep{2014ApJ...795..171S}, but as yet there is no confirmation of the scaling with magnetic field strength.

The work of \cite{2018MNRAS...submitted} elaborated the importance of the Alfv\'en crossing time ($\tau_A$) for the turbulent coronal heating process.
They found that increasing the loop length, which decreases $\tau_A$, leads to a decrease in the magnitude of the maximum electric current.
This implies that the turbulent heating process is becoming less efficient.
This conclusion was confirmed by \citep{2018ApJ...868.116D}, who found that when $\tau_p/\tau_A$ is under a critical value of about 4 (recall that
$\tau_p$ is the photospheric velocity time), the heating process will not produce enough radiation to permit, at present, any observation.
This implies that long coronal loops with small magnetic fields will not be observed, unless some other heating process is present.

In this paper we present a more detailed investigation of the properties of elementary events by analyzing the data of numerical simulations of the same loop at two different resolutions. Not all elementary events have the same geometrical size, the same duration and the same temperature peak and we present their distribution in size, duration and energy.  In Section 2 we summarize the mechanism responsible for the onset of the cascade of energy to small scales and for the formation of the current sheets. We describe how to relate such a dynamics to the definition of an elementary event.
In Section 3 we present the details of HYPERION and the initial and boundary conditions. In Section 4 Hyperion results in the case of $B_0 = 0.01$ Tesla and  $L_0 = 50,000$ km are analyzed. In Section 5 the statistical analysis performed on the data is used to derive the properties of the elementary events. We also present, for sake of comparison, some results at higher resolution to show the trend which justifies the choices made in the analysis at lower resolution. Section 6 is devoted to discussion and conclusions.

\section{Definition of Elementary Events}
\label{sect:definition}

The dynamics of the coronal loop are strongly influenced by the presence of a large-scale
mean magnetic field ($B_0$).
This field suppresses dynamics in the axial direction.
Disturbances propagate into the system from the boundary along $B_0$ and cause the growth of
perpendicular magnetic fields ($b_{\perp}$) and velocity fields ($v_{\perp})$
Dynamics in planes perpendicular to the axis then lead to the formation of electric current sheets.
Note that the ratio of the rms $b_{\perp}$ to $B_0$ is much less than one and is
also smaller than the aspect ratio.
Therefore an understanding of the energy dissipation can be obtained by studying a two-dimensional
section of the structure.
The evolution of the system, in the limit of a large loop aspect ratio, can be modeled by using the reduced MHD
equations (RMHD) \cite{Strauss 1976}.
The RMHD equations are valid for plasmas with a small ratio of
kinetic pressure to magnetic pressure, and in the limit of a small ratio of
perpendicular magnetic field to axial magnetic field ($b_{\perp}/B_0 \leq l_{\perp}/L_0$, where $L_0/l_{\perp} \gg1$ is the loop
aspect ratio).
Consequently, the typical velocities are also sub-Alfv\'enic.
In the RMHD model $b_{\perp}$ and $v_{\perp}$ depend on the axial coordinate ($z$).
Their nonlinear interaction proceeds independently on different constant $z$ surfaces along the loop.
Long-wavelength ($\lambda_A \simeq L$) Alfv\'en waves provide communication across axial planes.
These waves are the mechanism by which the energy originating from boundary motions propagates into
the interior of the system.

The nonlinear dynamics do not depend on the pattern of the velocity forcing chosen to mimic photospheric
motions, as appears evident by comparing the results obtained adopting very different photospheric velocity patterns. They have in common a photospheric velocity $u_p = 1$ km/sec, the spatial scale $l_p = 1000$ km and $\tau_A$ smaller than the photospheric convective timescale $\tau_p$.
\cite{2008ApJ...677.1348R} has adopted convection cell flow patterns constant in time, exciting all the wavenumbers $k$ and $l$ in the boundary planes included in the range $3 \le (k^2 + l^2)^{1/2} \le 4$. The results of the nonlinear dynamics induced by such forcing are very similar to those obtained by
\cite{2010ApJ...722...65R} where a photospheric velocity field in the form of a one-dimensional shear flow pattern is present. This case is interesting because it is possible to follow step by step the physical mechanisms responsible for the evolution of the system. The current layers in the perpendicular planes formed during the linear phase are unstable to tearing modes
\citep{Furth}, which are observed evolving during
the linear stage. Once the system becomes fully nonlinear the dynamics is fundamentally
different. The nonlinear terms do transport energy from the large to the small scales and the magnetic field topology does not maintain any resemblance to the tearing-like instabilities.

Reduced MHD studies have shown that
in the initial phase of development the perpendicular magnetic field grows linearly in time, obeying the relation $\mathbf{b}_\perp = \mathbf{u}_p t/\tau_A$.
During this phase $\mathbf{b}_\perp$ is a simple map of the velocity forcing pattern at the photospheric boundary.
The linear phase of growth continues until the perpendicular magnetic field attains a magnitude given by: ${b}_\perp \sim \ell_\perp B_0/L_0$.
This development occurs at a time given by $t_\ell \sim \tau_p > \tau_A$.
At this time a cascade of energy toward small scales develops.
The energy which is injected into the system at large spatial scales at the boundary is then balanced by the energy which is dissipated at small spatial scales in the interior of the system. The root mean square of $b_{\perp}$
and the Poynting flux also stop growing at this time,
both fluctuating around a mean value in the subsequent nonlinear phase of evolution.

It is important to note that, for a given $B_0$, the Poynting flux across the photospheric boundaries depends only on $b_{\perp}$ and the photospheric velocity.
The photospheric velocity is a boundary condition for the system the average magnitude of which is fixed.
The perpendicular magnetic field $b_{\perp}$, however, initially increases leading to a similar increase in the Poynting flux.
The Poynting flux then oscillates around a mean value in the subsequent non-linear phase.
Hence the energy available to form the current sheets that heat the loop is determined by the maximum value attained by $b_{\perp}$.
Since ${b}_\perp \sim \ell_\perp B_0/L_0$, this energy is ultimately dependent on the values of $L_0$ and $B_0$, the two significant parameters governing the perpendicular dynamics of the loop.
As the loop parameters vary, different regimes of turbulence can
develop.  Strong turbulence is found for weak axial magnetic fields
and long loops, leading to Kolmogorov-like spectra in the perpendicular
direction, whereas weaker regimes (steeper spectral
slopes of total energy) are found for strong axial magnetic
fields and short loops \citep{2007ApJ...657L..47R, 2008ApJ...677.1348R}.
Additionally they found that the scaling of the heating rate with $B_0$ is a power-law that
depends on the spectral index of total energy, and therefore its index changes with the
magnetic field intensity from $B_0^{3/2}$ for weak fields to $B_0^2$ for strong fields (for
loops with same aspect ratio).

Parker recognized that the transverse component of the magnetic field $b_{\perp}$ plays a key role in the evolution of the system.
In particular, he examined in great detail the inclination angle ($\theta$) of the mean $b_{\perp}$ to the mean field direction due to photospheric motions.
However, based on factors discussed earlier in this paper,
no relationship exists between the inclination angle ($\theta$) and the
rate of magnetic energy dissipation \citep[e.g., see][]{1996JGR...10113445G}.

The work of \cite{2015ApJ...submitted} and \cite{2018ApJ...868.116D} confirmed the dynamics described with the RMHD approach.
In addition, this research was able to determine how the coronal plasma heats
up and radiates energy as a consequence of photospheric convection
of magnetic foot points.
As previously stated, the loop is found to be a multi-thermal system in which there is a spatial correlation between the electric current
sheets and the temperature peaks.
High temperature regions are found to correspond to the locations of the electric current sheets, and extend axially along the loop magnetic field.
These isolated hot regions are surrounded by much larger regions of cooler plasma.
The sites of enhanced temperature represent the {\it {elementary events}}
defined above. They represent the building blocks of
what is called ``coronal heating'' which is actually the superposition of the radiation emitted by all
{\it {elementary events}}, averaged over time and spatial scales much longer than the typical dimensions and lifetime of a single elementary event. In the next Sections we will discuss the extent, duration and energetics of these elementary events.

\section{Simulation model}
\label{sect:model}
The HYPERION code is a computational representation of the corona.
In this representation the corona is modeled as a square cylinder (a figure which represents the computational box is discussed in Section 4) .
Along the $z$ direction a DC magnetic field is applied.
The boundaries in the $z$ direction represent the upper chromosphere.
At each $z$ boundary line-tied boundary conditions are enforced,
and the footpoints of this magnetic field are convected by applied flows.
Periodic boundary conditions are enforced in the $x$ and $y$ directions.
Hence the interior of the computational box represents the corona, while the $z$ boundaries represent the upper chromosphere.

\subsection{Governing equations}
As noted in the introduction, it is important to simulate all of the processes in the coronal energy cycle.
Magnetic reconnection is a significant process and hence the resistive terms must be included in the magnetic induction equation to ensure that it occurs correctly.
The Joule heating term is needed in the energy equation to ensure that the thermal energy created by magnetic reconnection is deposited in the correct location and in the proper amount.
Thermal conduction and optically thin radiation are needed to obtain the effects of the chromospheric response to heating, such as evaporation.
In addition, compressibility must be factored in to allow for stratification effects due to gravitation.
Hence the equations used in the simulation model should allow for all of these effects.
The HYPERION code evolves in time the three-dimensional compressible magnetohydrodynamic equations.
Viscous and resistive dissipation terms are included, as well as temperature dependent thermal conduction and optically thin radiation.
The governing equations are as follows:
\begin{eqnarray}
{{\partial n}\over {\partial t}} &=& -\nabla\cdot (n {\bf v}),  \label{eq:eqn}      \\[.4em]
{{\partial n {\bf v}}\over{\partial t}} &=& -\nabla\cdot({n \bf v v})
   -{\beta}\nabla p + {\bf J}\times{\bf B}
 + {1\over S_v}\nabla\cdot{\bf \zeta}\\ \nonumber
&& + \frac{1}{Fr^2}\   n \Gamma(z)\, {\bf\hat e}_z \label{eq:eqnv} \\[.4em]
{{\partial T}\over{\partial t}} &=& -{\bf v}\cdot\nabla T
 - (\gamma - 1) (\nabla\cdot {\bf v}) T \\
\nonumber
&& +\frac{1}{n} \Bigg\{ \frac{1}{ Pr\, S_v }
\bigg[{\bf B}\cdot\nabla
\bigg( \kappa_{\parallel}\ T^{5/2}\ {{\bf B}\cdot\nabla T\over B^2}\bigg) \bigg]
\nonumber \\
&& +{(\gamma -1)\over\beta}  \bigg[
{ 1\over S_v} \zeta_{ij} {\partial v_i\over\partial x_j}
+{1\over S} (\nabla\times{\bf B})^2 \\
&& -{1\over P_{rad} S_v} n^2\Lambda (T)
+ {\beta\over(\gamma - 1)} n C_N \bigg] \Bigg\}, \label{eq:eqT}\\[.4em]
{{\partial {\bf B}}\over{\partial t}} &=& \nabla\times{\bf v}\times{\bf B}
  - \frac{1}{S}\nabla\times \nabla\times {\bf B}, \label{eq:b}\\[.4em]
%\end{eqnarray}
&&\nabla\cdot{\bf B} = 0, \label{eq:divb}\\[.4em]
\nonumber
\end{eqnarray}
\noindent
\begin{equation}
p = nT. \label{eq:eqp}
 \end{equation}

As is typical in turbulence calculations, these equations are written in
dimensionless form.
The dimensional terms are lumped together to form dimensionless numbers.
The dimensionless numbers are:
the viscous Lundquist number :$S_v = n_* m_p  V_{A*} L_* / \mu$
($m_p = 1.673\times 10^{-27}$ kg is the proton mass),
the (resistive) Lundquist number: $S = \mu_0 V_{A*} L_* / \eta$
($\mu_0 = 1.256\times10^{-6}$ Henrys / meter is the magnetic permeability),
$\beta = \mu_0 p_* / B_*^2 \equiv$ boundary pressure ratio,
The Prandtl number: $Pr = C_v \mu / \kappa_{\parallel} T_*^{5/2}$,
the radiative Prandtl number: $P_{rad} $ where
${\mu/ \tau_A^{2} n_*^2 \Lambda (T_*)} $. $C_v$ is
the specific heat  at constant volume, and
the magnetohydrodynamic Froude number ($Fr = V_A/(g L_*)^{1/2}$,
where the solar surface gravity is given by: $g=274$~m~s$^{-2}$.

The non-dimensional variables are defined in the following way:
the number density is $n ({\bf x}, t)$,
the flow velocity is ${\bf v}({\bf x}, t) = (u, v, w)$,
the thermal pressure is $p({\bf x}, t)$,
the magnetic induction field is ${\bf B}({\bf x}, t) = (B_x, B_y, B_z) $,
the electric current density id ${\bf J} = \nabla\times{\bf B}$,
and the plasma temperature is $T({\bf x}, t)$,
Also note that the viscous stress tensor is $\zeta_{ij}= \mu (\partial_j v_i + \partial_i v_j) -
\lambda \nabla\cdot {\bf v} \delta_{ij}$,
the st4rain tensor is $e_{ij}= (\partial_j v_i + \partial_i v_j)$,
and the adiabatic ratio is $\gamma$.

The characteristic values of various quantities are chosen to represent
values typical of the upper chromosphere.
These characteristic values are then used to render the equations dimensionless.
These values include the characteristic number density ($n_*$),
the characteristic temperature ($T_*$),
and the characteristic parallel Alfv\'en speed [$V_{A*}$])
The perpendicular box width ($L_*$) is also used (here chosen to be $4\times 10^6$ m).
Time ($t$) can then be measured in units of the dimensionless Alfv\'en time $\tau_A$,
which is given by $\tau_A=L_* /V_{A*}$.

The shear viscosity and magnetic resistivity ($\mu$ and $\eta$) are
spatially uniform and temporally constant.
The bulk viscosity ($\lambda$) is set to equal to $(2/3) \mu$, {\it i.e.,}
Stokes relationship is enforced.
The term $\kappa_\parallel$ denotes the parallel thermal conductivity.
$\kappa_{\perp}$, the perpendicular thermal conductivity, is neglected.
The CHIANTI atomic database \citep{1997AAS...125...149,2012ApJ...744...99L} is used to
construct the radiation function.
The base temperature of the loop ($T_* = 20000 K$) is used to normalize the
radiation function.
An elliptical model is used to construct the loop gravity profile $[\Gamma (z)]$.
Near the bases of the loop Newton cooling ($C_N$) is effective \citep{2001A&A...365..562D}.
Let $T_i (z)$ denote the initial temperature profile,
at the upper boundary we enforce $C_N = 10~[T_i (z) - T(z)] e^{-4(0.5L_z-z)}$
while we set
$C_N = 10~[T_i (z) - T(z)] e^{-4(z+0.5L_z)}$ at the lower boundary .

The numerical method used to solve employs Fourier collocation-finite difference scheme in space.  Time step splitting  is employed in time.  Super TimeStepping is used for thermal conduction \citep{2012MNRAS.422.2102M}.  All other terms are advanced in time with a low-storage fourth-order Runge-Kutta scheme \citep{carpenter1994fourth}.
More detail on the numerics also can be found in our previous papers
\citep{2015ApJ...submitted, Dahlburg2016b,2018ApJ...868.116D}.

\subsection{Initial and boundary conditions} \label{sec:inibc}

Let $\mathbf{A}$ be the vector
potential associated with the fluctuating magnetic field.
Then the total magnetic field in the loop is given by
$\mathbf{B} = B_0 \mathbf{\hat e}_z + \mathbf{b}$
with $\mathbf{b} (x,y,z,t)=  \nabla \times \mathbf{A}$.
Initially the fluctuating magnetic field and the velocity field are set to zero in the loop,
At the $z$ boundaries the magnetic field ($B_z$), the number density ($n$) and the temperature ($T$)
are kept constant at their initial values, which are, respectively, $B_0$, $n_0$ and $T_0$,
while the perpendicular magnetic vector potential ($A_x$ and $A_z$ are convected by the flows due to the boundary forcing function.
The initial profiles of $n$ and $T$ are determined by an elliptical gravity model
a full description of which is given in the next subsection.

The forcing function, first adopted by \cite{1996ApJ...457L.113E}, is designed to inject energy into this system within a long-wavelength
forcing band and also to permit the generation of both harmonic and subharmonic disturbances.
The dominant length of the forcing is approximately 1000 km in dimensioned form.
The dimensionless forcing time is set to 16.3548, a value that represents a physical time of 300 seconds.
Hence the forcing represents the energetics of a photospheric granule.
A full discussion of the initial conditions and boundary conditions is found in \cite{2015ApJ...submitted}.

\subsection{Initial temperature, number density and gravity specification} \label{sec:grav}

Heating is a consequence of magnetic reconnection in elementary events.
Starting from an initial uniform potential magnetic field ($B_0$),
it takes some time to build up a large enough perpendicular magnetic field to permit this to happen.
If thermal conduction and radiation are turned on from the beginning of the simulation, the initial
temperature profile is significantly altered.
In the absence of a heating function an energy balanced equilibrium cannot be specified.
Hence we begin with an ideal thermodynamic equilibrium and ramp up the thermal conduction while
the magnetic stresses build up.
Here the time to ramp up from zero to the full value is chosen to be five minutes.
The initial temperature profile ($T_i$) is set to $2\times10^4$ K at the boundaries and $10^6$ K in the center
(for the nondimensional values see \cite{2015ApJ...submitted}).

\begin{figure}
\centering
 \includegraphics[width=0.9\columnwidth, bb = 150 40 635 500]{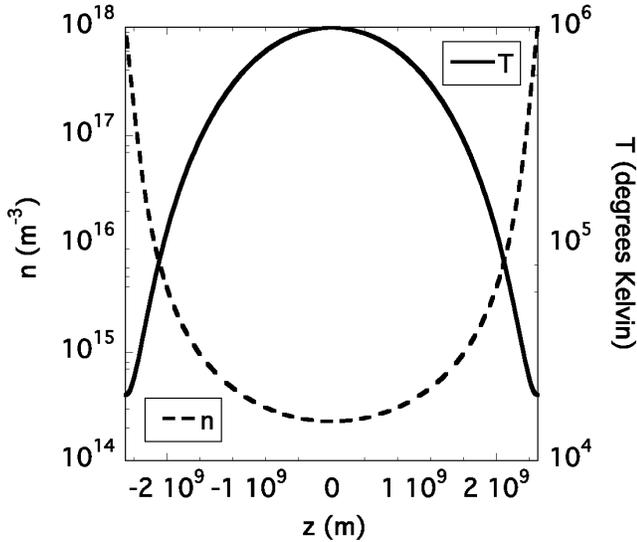}
 \caption{\label{rtinit} The initial number density ($n$) and temperature ($T$)  {\it vs} $z$ used in both numerical
  simulations.\\}
\end{figure}

The loop gravity is given by an elliptical model which is described in \cite{2015ApJ...submitted}.
With the initial temperature and gravity specified, the corresponding number density can be
found numerically with a shooting method, as described in \cite{2015ApJ...submitted}.
In this paper we use a dimensionless loop length consistent with a dimensional loop length of 50000 km.
We set a dimensionless value of one for the number density at the loop boundary.
The shooting method then give a dimensionless number density value of $2.298\times10^{-4}$ at the loop apex.
In dimensioned terms, the initial number density is $1.0\times10^{18}$ m$^{-3}$ at the loop boundary
and $2.298\times10^{14}$ m$^{-3}$ at the loop apex.
The profiles of the initial number density and temperature used in both numerical simulations are
shown in Figure \ref{rtinit}, in which they are restored to dimensional form.

\subsection{Selection of parameters}

In this paper we examine a loop of length of $L_0$= 50,000 km and a magnetic field strength of $B_0=0.01$ Teslas. For the number density, we set $n_0=10^{18}\,$ m$^{-3}$, and for the temperature $T_0=2\times10^{4}\,$  K, and $L_0 = 4 \times 10^{6}\,$  m. The Coulomb logarithm
($\ln \Lambda$) is set at a value of 10. A five minute convection time is used to determine the normalized forcing time scale, $t^*$.
In this paper two numerical simulations are described.
The higher resolution simulation has less physical dissipation than the lower
resolution simulation.
The first simulation, which is run at a resolution of $128\times128\times192$ has
$S_v=1.745\times10^5$,
$S=2.882\times10^6$,
$Pr=1.695\times10^2$, and
$P_{rad}=1.337\times10^{-18}$,
where
$S_v$ is the viscous Lundquist number,
$S$ is the resistive Lundquist number,
$Pr$ is the Prandtl number, and
$P_{rad}$ is the radiative Prandtl number.
The second simulation, which is run at a resolution of $ 256\times 256\times192$ has
$S_v=3.490\times10^5$,
$S=5.763\times10^6$,
$Pr=8.475\times10^1$, and
$P_{rad}=6.683\times10^{-19}$.
Both numerical simulations have $\beta = 3.471\times10^{-3}$ and $Fr=6.587$.
Numerical resources prevent the simulation of solar Lundquist numbers and Prandtl numbers.
The Lundquist numbers are reduced to allow for proper numerical resolution.
The Prandtl numbers are rescaled so that the thermal conduction has the same  efficiency relative to the dissipation in the energy equation that it has in in the solar corona.
An extended discussion of this point is found in \cite{2015ApJ...submitted}.

\section{EVENTS IDENTIFICATION}
\label{sect:events}
In this section we describe the algorithms that we developed to identify elementary
events in the HYPERION simulation runs. Given our experience studying the
dynamics generated by footpoint driven motions that braid the field forming elongated
current-sheets that are followed by a sudden release of energy and increases in
temperature, we expect heating events to be transient and localized changes in the
MHD properties of the plasma.

In Section~\ref{sect:definition} we defined elementary events as the temperature
enhancements associated to current sheet formation. Temperature increase is just
a consequence of the heating, and therefore this definition does not allow us to
call them elementary heating events from the outset. We will argue through this
paper, however, that elementary events identified in temperature space can be a
good proxy for certain properties of the elementary heating events, that otherwise
are better described by terms such as the Joule heating.

There are several reasons to choose temperature over other plasma properties to
identify elementary events in the corona: 1) the temperature increase is directly
connected to the heating exchange that we are trying to investigate; 2) the
temperature response of the plasma is also tied to the radiative response and
therefore observational diagnostics; 3) the temperature maps are smooth and easy to be
interpreted by the algorithms we have developed. Density is important in the
radiative response, but it is not as closely connected to heating because of its
slower response \citep[e.g.][]{2014LRSP...11....4R}.
For completeness we describe at the end of this section the results from identifying events as transients of the Joule heating term, and discuss the similitudes and differences with the temperature inferred results. While the Joule heating term, $\eta J^{2}$, directly measures the energy release, this term cannot be measured observationally. The relationship between the current and radiative emission is not straightforward. Furthermore, identifying events using this term presents challenges to the performance of the identification algorithms, that make it best to address it after the temperature case.

Identifying events in space and time within the simulation box involves locating
neighboring pixels with a shared evolutionary history for a relevant property.
Machine learning, and particularly clustering algorithms, have been a popular and
efficient technique used to address this problem in a variety of contexts, including
Solar Physics \citep[]{caballero2013}, where automated data processing methods have
blossomed in the past decade with the high data volumes of the Solar Dynamics Observatory
\citep[e.g.][]{aschwanden2010,martens2012,bobra2018}.
We describe below two clustering approaches that we have followed in assessing our ability
to uniquely identify and characterize elementary events in space and time. We present both to show
the robustness of the results. The first algorithm uses a well known clustering
technique called density-based clustering of applications with noise (DBSCAN). The
second is our own developed brute force cluster finder that relies in a distance
search instead of the pixel search that characterizes our DBSCAN version. While the working
philosophy is similar, definitions and search strategy make them distinct testbeds
for this experiment, and the comparison gives us some insight about the uncertainties
of the process and the results.

\subsection{Density Based Clustering (DBSCAN)}
\label{subsec:dbscan}
The density-based clustering of applications with noise (DBSCAN) is a data clustering
technique proposed by \citet{ester1996}. DBSCAN finds clusters of points within a
sample with the assumption that a minimum number of neighbors are within a distance
radius of a given point in the cluster, i.e. the density of points in the neighborhood
is higher than a certain threshold.

\begin{figure*}
\centering
\includegraphics[width=0.33\linewidth]{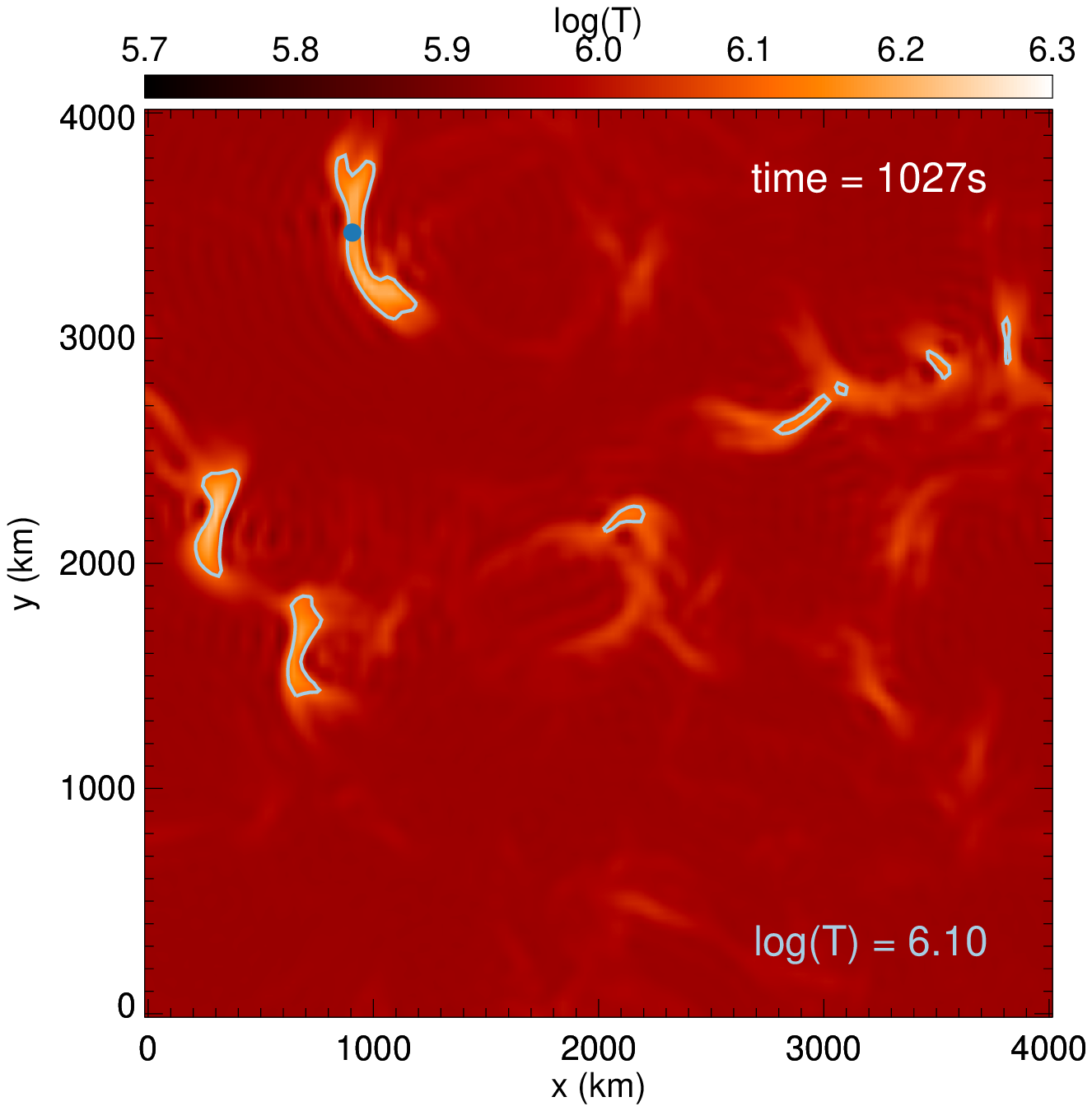}
\includegraphics[width=0.33\linewidth]{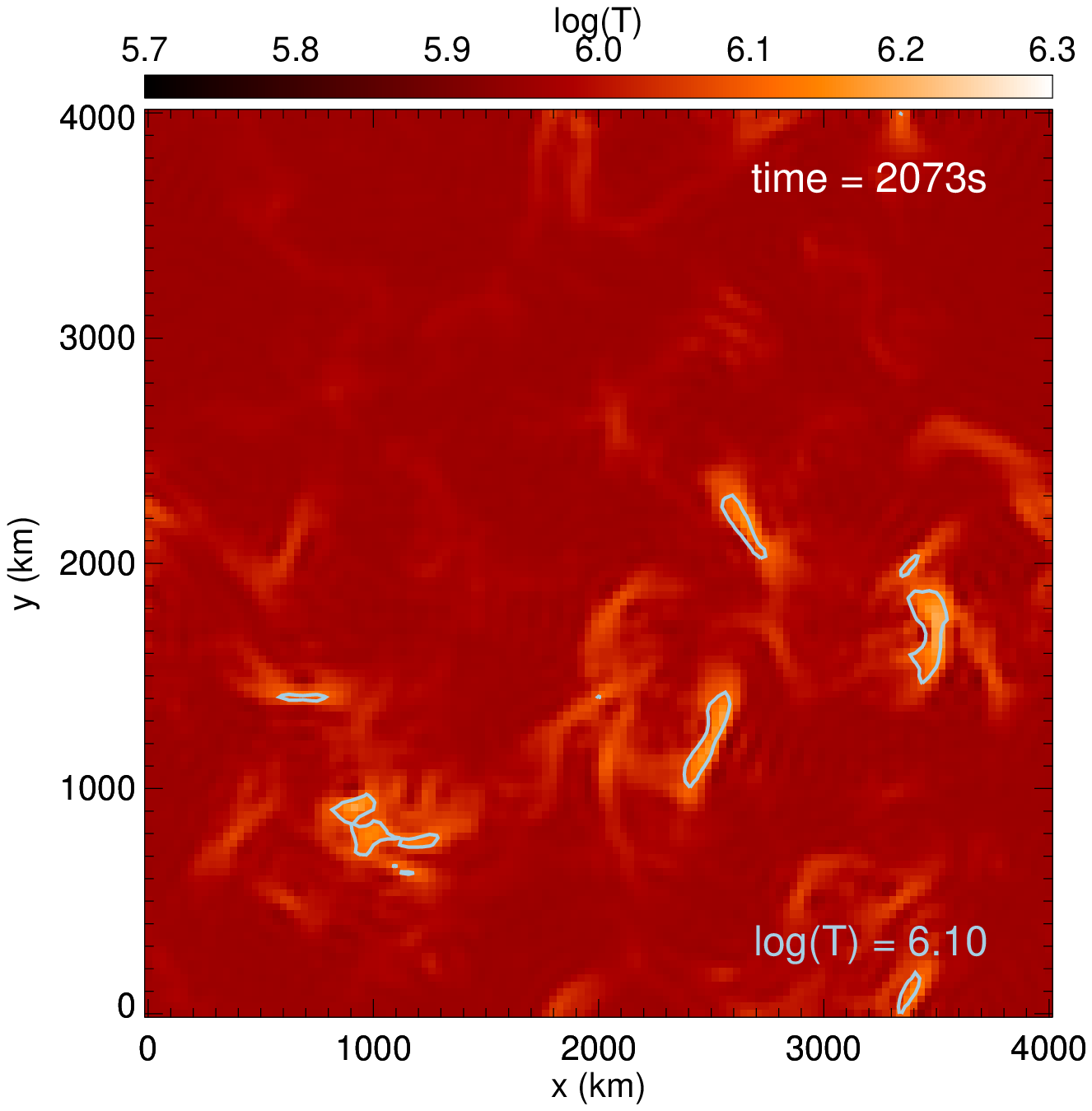}
\includegraphics[width=0.33\linewidth]{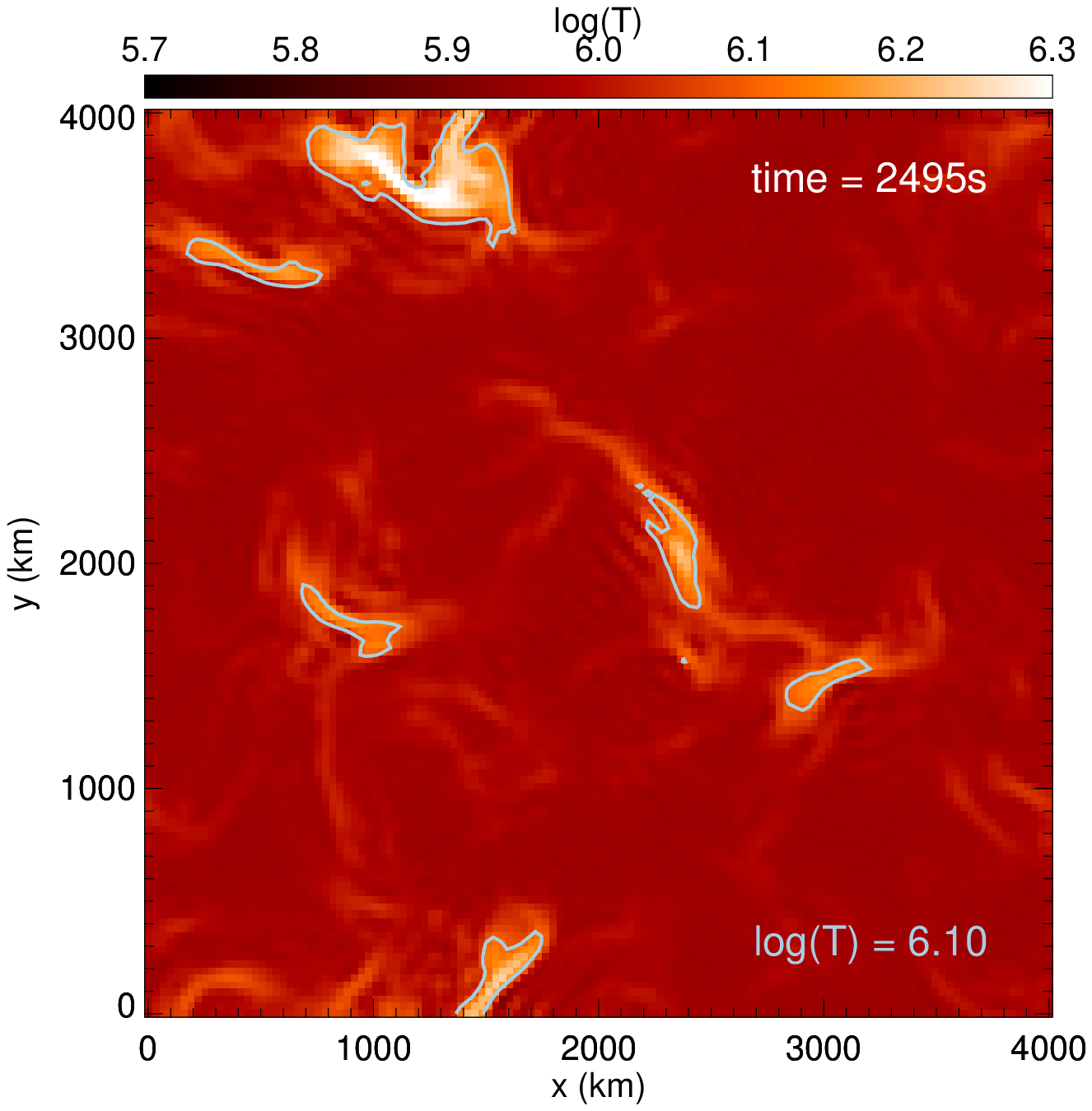}
\caption{\label{fig:temperature} Temperature distribution at the middle cross-section,
apex of the loop, for three time steps. The contours highlight the temperature
threshold used to identify heating events. The blue circle highlights a location
in a cluster that is referenced in Figure~\ref{fig:tempprof}.}
\end{figure*}

We have developed an Interactive Data Language (IDL) algorithm that applies this
technique to any $n\times m$ array, where $n$ is the number of variables and $m$
the number of observations, using as reference the definitions, assumptions
and the algorithm description in the \citet{ester1996} paper and an existing
Python implementation\footnote{\tt{https://github.com/chrisjmccormick/dbscan/}}.
The search evaluates the number of points in the sample that are within a predefined
neighborhood radius $\epsilon$ (in pixels) of a given point and makes decisions
about whether that point is part of a new or existing cluster.
If the number is equal or larger than the required threshold ($N_{pts}$) then all
the identified points are considered part of the cluster. Core points in the cluster
are those that satisfy this condition about their neighborhood. Border points are
those that have less than $N_{pts}$ in their neighborhood, but belong to the cluster
because of the core point they are connected to. The algorithm evaluates the
neighborhood of every point in the sample and stops when all have been classified
as cluster or noise. Noise points do not have any core cluster neighbors in their
$\epsilon$. For the purpose of our study, noise points are discarded.

In our case $n=3$, the three spatial dimensions of the simulation box in pixel units, and $m$ is the number of points per time step that we aim to classify. As
stated before, we make identifications based on the maximum temperature. We use a hard threshold ($\tau$) to locate and extract those $m$ locations and then
apply DBSCAN to return the number of clusters within that set, and the unique label for all points in each cluster. This is performed for all time steps. The results
that we present here are made with $\epsilon=1.5$, $N_{pts}=2$. We experimented with various temperature thresholds and settled with $\tau_{logT}=6.10$.
That temperature captures temperature events that are visually discernible in a cross-section at the loop apex, keeping the number of points and clusters in a
manageable range to perform the temporal tracking and to compute their properties.
This definition makes it more likely to detect events at the apex, because the
apex of the loop generally has the highest (coronal) temperature. Additionally, the apex
typically has the lowest density, so events occurring at this location will in general
release more energy per particle, and thus cause the largest temperature deviations.
We should note that other temperature event definitions are possible. For instance,
one that instead of focusing on coronal events, looks for departures of temperature from
a steady state.
Figure~\ref{fig:temperature} shows the cross-section and reference thresholds for three time steps.

\begin{figure*}
\centering
\includegraphics[width=\linewidth]{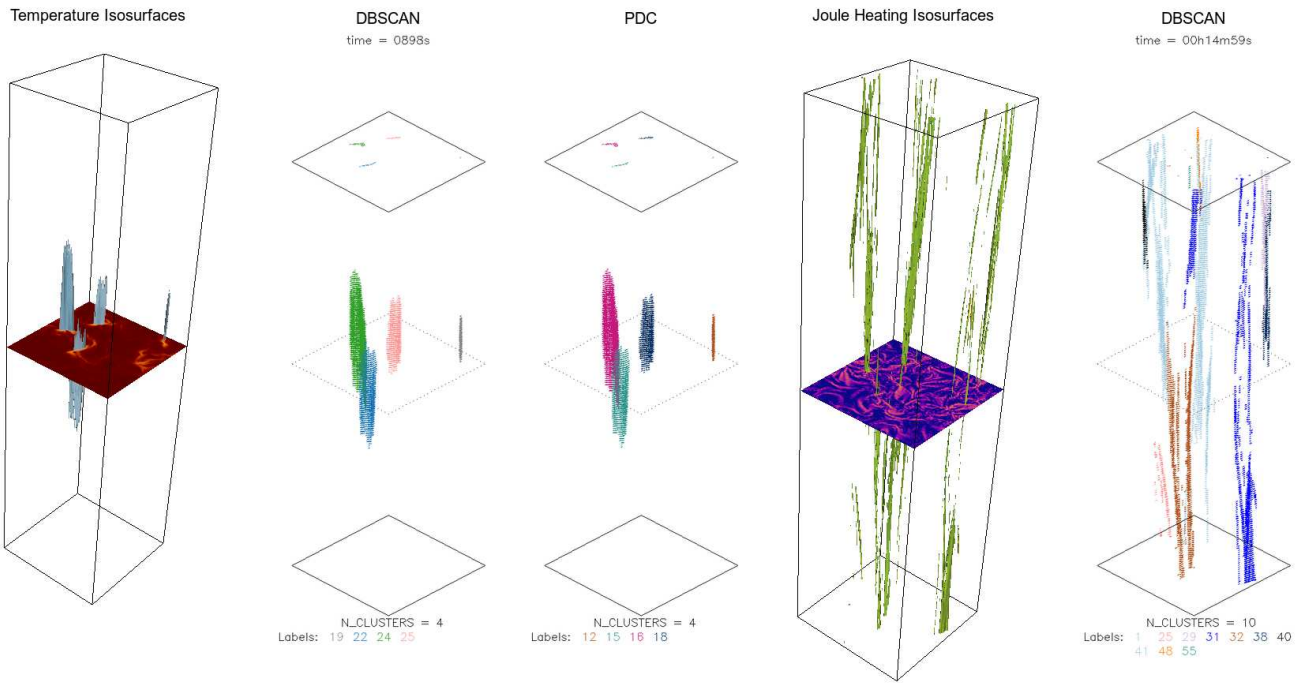}
\caption{\label{fig:events} Heating events identification for one of the time steps in the
HYPERION run compared to the isosurfaces of a given plasma property. The left panel shows
temperature isosurfaces ($log(T)=6.10$) in blue and a XY slice of the temperature
distribution perpendicular to the Z direction at the loop's apex. The next two panels
show the events (clusters) identified by the DBSCAN and PDC algorithms. Different colors
correspond to different cluster labels. The right panels show the Joule Heating isosurfaces
($1.0\times 10^{-5} \,\rm Watts\,m^{-3}$) in green with the corresponding XY slice cross-section and the DBSCAN
identifications. A movie version is available online. The movie starts at time 183s and
ends at 3100s and it shows the evolution of the temperature and Joule Heating
isosurfaces as a function of time, as well as the event identifications from the two methods
at every time step. The current figure is one of the frames in the movie (time = 898s).}
\end{figure*}

Once we have identified all clusters of points, i.e. our definition of elementary
events, in  all time steps, we apply a second  algorithm that we call here ``cluster tracker" that determines
if clusters from consecutive time steps are the same and should have the same label.
The algorithm was also developed in IDL and tracks the events forward in time, by
determining whether there is a spatial overlap between the clusters in consecutive time
steps. Events that appear as a single cluster in a given time step, can then split
in multiple fragments as they evolve in time and retain a unique label, but events
that form from coalescing clusters keep their independence before the merger. The
cluster with the larger size dictates the label for the cluster going forward.

Figure~\ref{fig:events} shows on the two left-most panels the temperature
isosurfaces at the $\tau_{logT}=6.10$, together with the event identification.
The algorithm correctly identifies four data clusters corresponding to four
heating events. A movie version of the figure is available online. The movie goes with the paper.

\subsection{Physical Distance Clustering (PDC)}

Since distinct clustering methods can produce distinct results on the same data set (see, for example, the description of clustering methods in the Python package Scikit-learn \citealt{pedregosa2011}), we have implemented a second method in IDL to test the consistency of the results.  In the DBSCAN method described in the previous section, clusters are calculated by searching over distances measured in pixels and finding clusters above a certain density.  However, in HYPERION, the voxels are not cubes but are elongated in the vertical direction $z$, that is, $dz > dx$ and $dx = dy$.
In the first simulation, examined here, $dz \approx 26$\,km and $dx = dy = 3.125$ km. We therefore have additionally implemented a clustering method that searches based on physical distance (physical distance clustering, PDC) rather than pixel distance.  In principle, the two methods should give similar results and so we can compare each method.

The code works by first identifying all voxels in the grid where a variable such as temperature is above some input threshold.  For example, it first locates all voxels above 1 MK.  Then, it iterates over that subset of voxels to determine which other voxels are within a radius of defined distance, which are then labeled as belonging to the same cluster of points.  In the results presented in Section \ref{sec:analysis}, this is defined as a distance of $2\ dz$.  Note that since $dz > dx$, this implies that the radius in number of pixels is larger in $x$ and $y$ than in $z$.  The code then iterates over the clusters to make sure any overlapping clusters have the same label.  Finally, the code iterates over each time snapshot, checking for positional overlaps between clusters at two adjacent times.  If there is an overlap, the clusters are assumed to be the same event and labeled accordingly.

This method then gives us a list of clusters, along with their locations at each time, and we store the parameters of each voxel within each cluster.  Using this, we can then easily calculate the internal energy, size, duration, aspect ratio, etc., of the clusters to build up statistical information about the events.  In Figure \ref{fig:events}, the clustering results in a HYPERION snapshot are shown in the third panel, for comparison with DBSCAN in the second panel.  At many times, the methods find the same results, but in general, PDC locates fewer events because it clusters together many smaller events than DBSCAN does.  This difference can be seen in the statistical analysis presented in the next section.

\section{Data analysis}
\label{sec:analysis}

Being able to identify and track clusters of points in time allows us to determine
the properties of the events as they evolve in the simulation run.
As the HYPERION numerical solution provides the changes in temperature, density,
velocity, current, etc. for every one of the pixels, we can compute the characteristic
properties of the set of pixels that conform every event in our identification.
In this section we present the results from this analysis. First, we present the
properties of heating events identified in the temperature domain. Then, we discuss
how the results compare when looking into the Joule heating space.

\begin{figure}
\centering
\includegraphics[width=\linewidth]{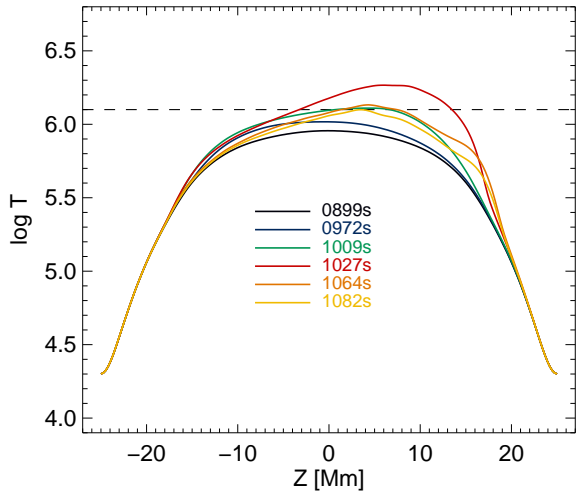}
\caption{\label{fig:tempprof} Temperature evolution in time along the Z direction
at the location of the heating event highlighted with a blue circle in Figure~\ref{fig:temperature}.
The dashed line shows the threshold level used in the event identification.}
\end{figure}

\subsection{Temperature events}
Figure~\ref{fig:tempprof} shows the temperature changes in time along the Z
direction at the location of a temperature enhancement highlighted
in Figure~\ref{fig:temperature} with a blue circle. This is the first event at that
location in this run and the pre-event temperature profile is very similar
to the profile at the start of the run. The heating resulting from the forcing at the
footpoints produces a temperature spike of approximately 1.55 MK, offset towards the
top boundary of the box (positive Z). This temperature profile is not the temperature
distribution along an isolated field line, but it is representative of the general
temperature changes observed along the box, i.e. along the loop, when a heating event
takes place. The figure also shows
the threshold level that sets the definition of a heating event in this analysis. Only
voxels with temperatures above that threshold survive the identification process and
temporal tracking depicted in Figure~\ref{fig:events}. During the evolution of the event
the density exhibits small departures from its initial profile.

\begin{figure*}
\centering
\includegraphics[width=\linewidth]{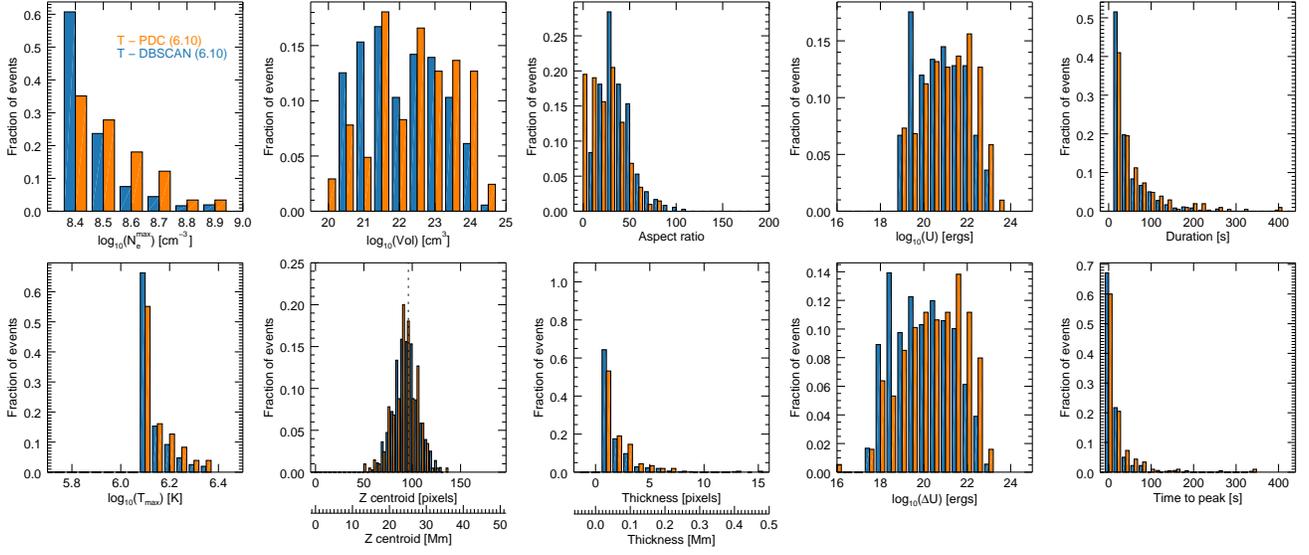}
\caption{\label{fig:comparison_methods} Properties of the heating events identified
in the temperature domain by the two clustering algorithms. DBSCAN and PDC identified
359 and 205 events respectively. The dotted line in the Z centroid panel represents
the midpoint distance along the flux tube.}
\end{figure*}

\begin{figure*}
\centering
\includegraphics[width=\linewidth]{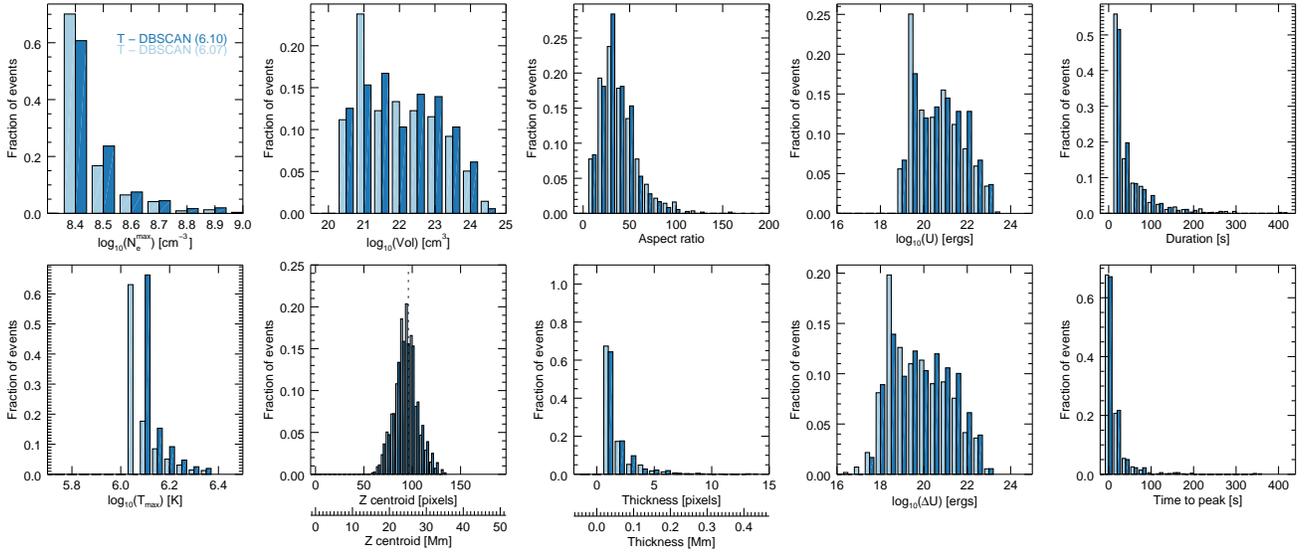}
\caption{\label{fig:comparison_thresholds} Properties of the heating events identified
by DBSCAN using two different temperature thresholds:
 $\tau_{logT}=6.10$ and  $\tau_{logT}=6.07$. The number of events identified are 359
 and 555 respectively}.
\end{figure*}

For every identified event we computed the total volume as the sum of all voxels,
the extent in the three spatial dimensions, the duration, the centroid position,
the maximum temperature and density and several other derived quantities, such as
aspect ratio, thickness or internal energy. These quantities were calculated for
every instant in the lifetime of an event. In Figure~\ref{fig:comparison_methods}
we show the summary of these properties for the events identified by the two
clustering algorithms. We present these results in CGS units, as opposed to the SI units of previous figures, to facilitate comparisons to observables that are usually reported in those units. In the case of volume, aspect ratio, centroid, thickness
and energies the values correspond to the time step of maximum temperature in each
cluster, which marks the end of the heating. The purpose of showing the comparison
of both algorithms is to illustrate the differences that the choice of algorithm
has on the results. While DBSCAN and PDC identified a different number of events,
359 and 205 respectively, the overall distributions are very similar. The larger
differences are in volume, as PDC sometimes clusters together events separated by
DBSCAN, and internal energy which is highly correlated to volume.

The left column shows the distribution in the maximum density and temperature reached
by any point within the cluster. The distribution is dominated by low temperature
and low density events, close to the temperature threshold, suggesting that as we
lower that threshold the number of events will increase and their peak temperature
will go down. Figure~\ref{fig:temperature} already hints that we are missing events with
our choice of threshold. To verify this intuition we run both algorithms with a
threshold of $\tau_{logT}=6.07$ and the trend was confirmed as shown in
Figure~\ref{fig:comparison_thresholds} for DBSCAN. The number of events increased to 555.
We had difficulties working with lower thresholds because the increased number of
points creates challenges to the 'cluster tracking' part of the workflow.
We are working in making the algorithms more efficient for future studies.

One of the most fundamental quantities of the heating events that we want to know
about is energy. In Figure~\ref{fig:comparison_methods} we show the maximum
internal energy ($U=\kappa_B N_e T_e$) reached by all clusters, computed as the
sum of all voxels. This quantity, however, is only indicative of the energy in
the system at this moment in time. It does not tell us what was gained with the
most recent heating event. For that we calculate an internal energy increment
from the difference in internal energy between the energy at peak temperature and
the start of the event:
\begin{equation}
\Delta U=\sum_{ijz}\kappa_B N_e T_e \Biggr\rvert_{t(T_{max})} - \sum_{ijz}\kappa_B N_e T_e \Biggr\rvert_{t = 0}
\end{equation}
where $i$,$j$,$z$ are the spatial indices of the voxels that define the cluster at maximum temperature. This measure of the Joule heating is strictly less than the total energy release.  This is because some of the energy is radiated away, conducted away, or advected away from the site of release.  However, we argue that despite this limitation it is a useful proxy for the energy release because the bulk of the released energy goes into heating the plasma in the loop. In Section~\ref{sect:joule} we discuss further the validity of this assumption. In Section~\ref{sect:kolmogorov} we discuss the shape of the energy distributions.

Figure~\ref{fig:comparison_methods} shows the distribution of the clusters centroid positions along the loop (Z centroids), which tells us that the temperature events are distributed in both directions around the loop apex, but very close to it. This is not surprising given our choice of temperature as a proxy for heating. Coronal temperatures are more likely to be detected at the apex. As there can be differences between where the temperature peaks and where the heating is deposited, we examine identifying events with the current distribution in the next section. We also computed the thickness of the structures at the apex plane. The thickness is defined here as the minimum value of the median thicknesses in X and Y. This preserves the smaller dimension of the two in an elongated (e.g. elliptical) cross-section. The duration of the events is shown on the right most column and reveals that most events are less than 100\,s long, and about half that time for the time to reach the maximum temperature.

\subsection{Joule heating events}
\label{sect:joule}

\begin{figure}
\centering
\includegraphics[trim=5cm 3cm 5cm 4cm,clip,width=\linewidth]{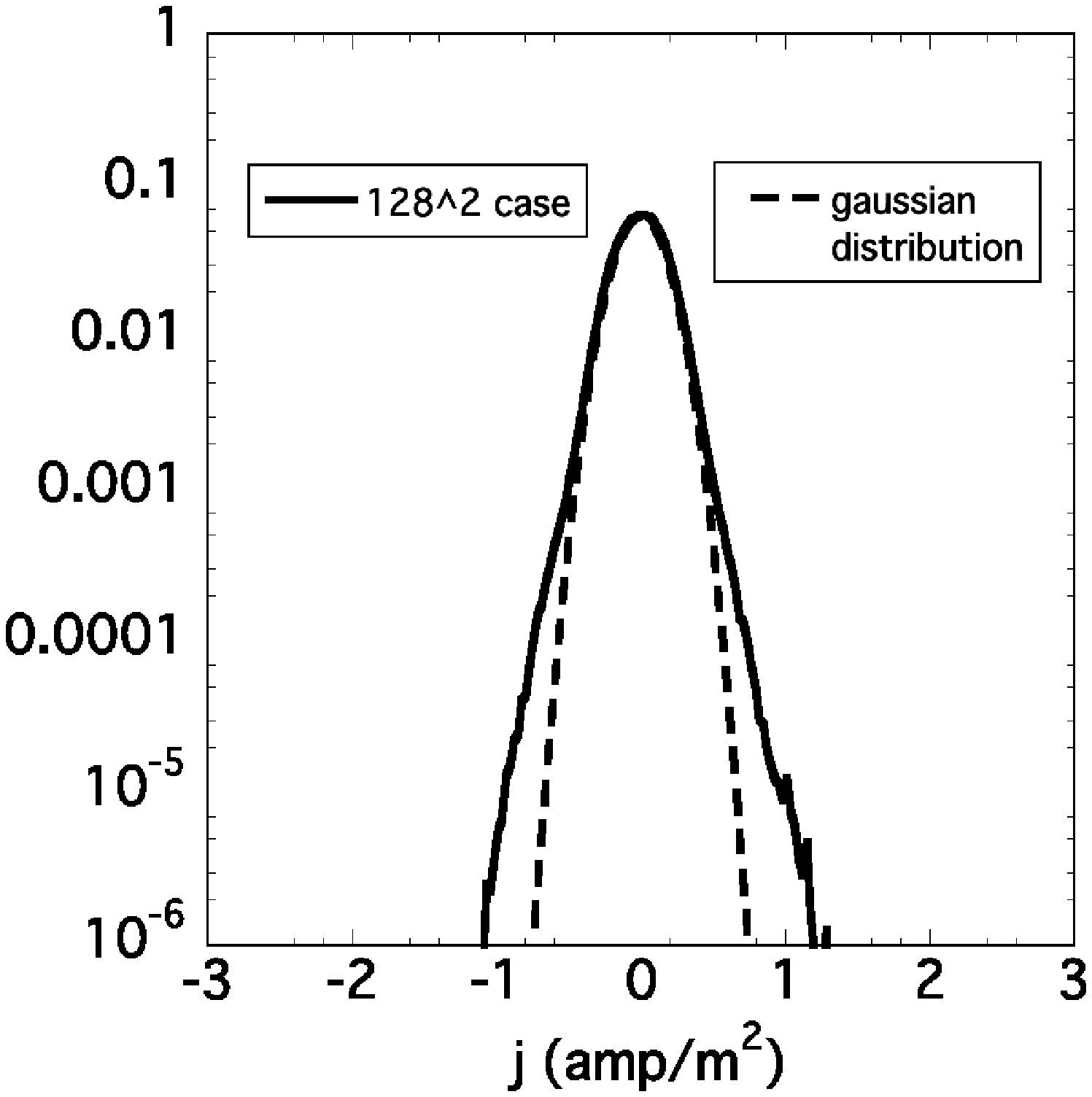}
\includegraphics[trim=5cm 1cm 5cm 4cm,clip,width=\linewidth]{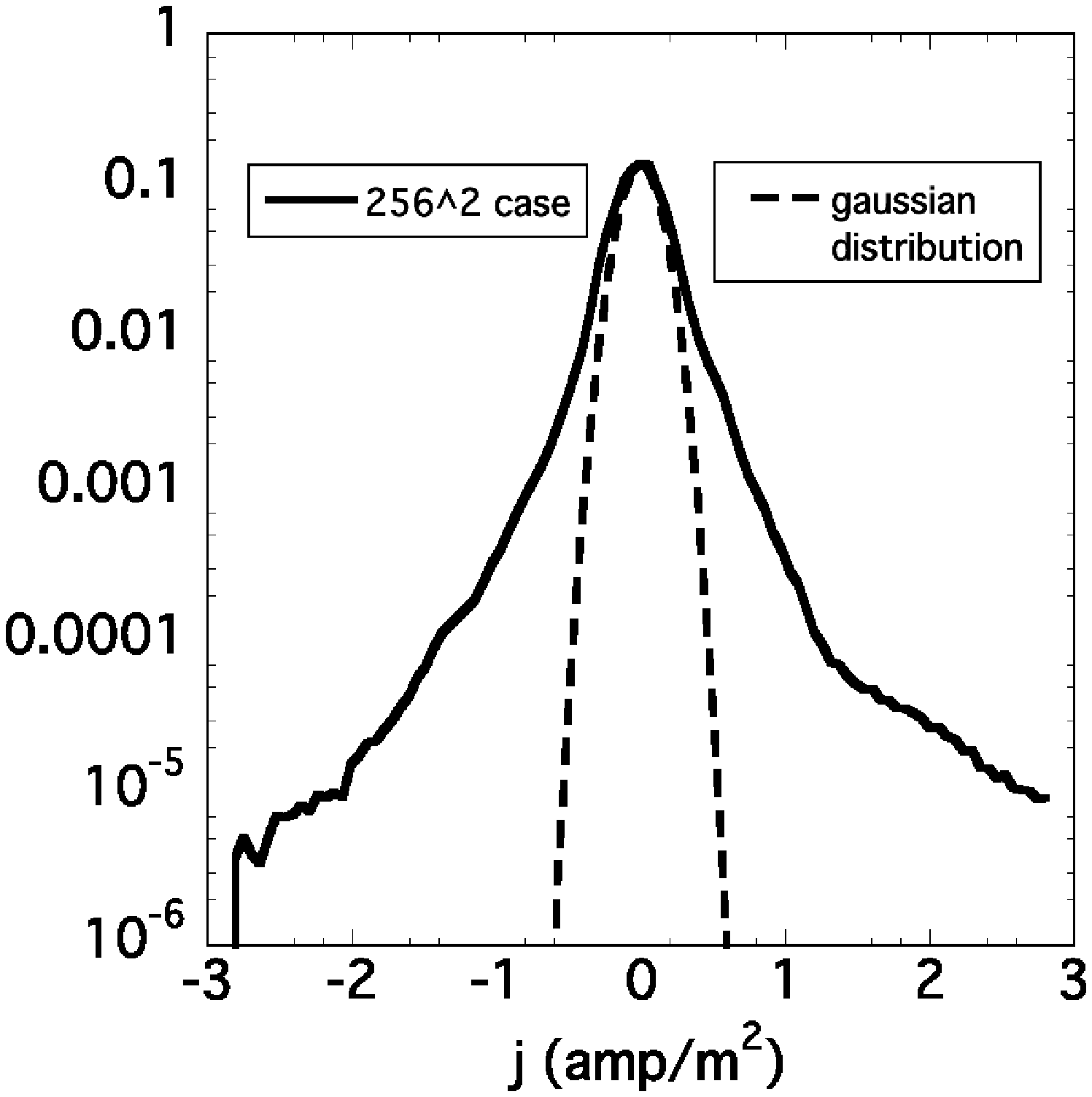}
\caption{\label{fig:PDF} Probability Density Functions for the electric current distribution inside the box for a time step in the low resolution Hyperion run (top) and the high resolution (bottom). }
\end{figure}

\begin{figure}
\centering
\includegraphics[width=\linewidth]{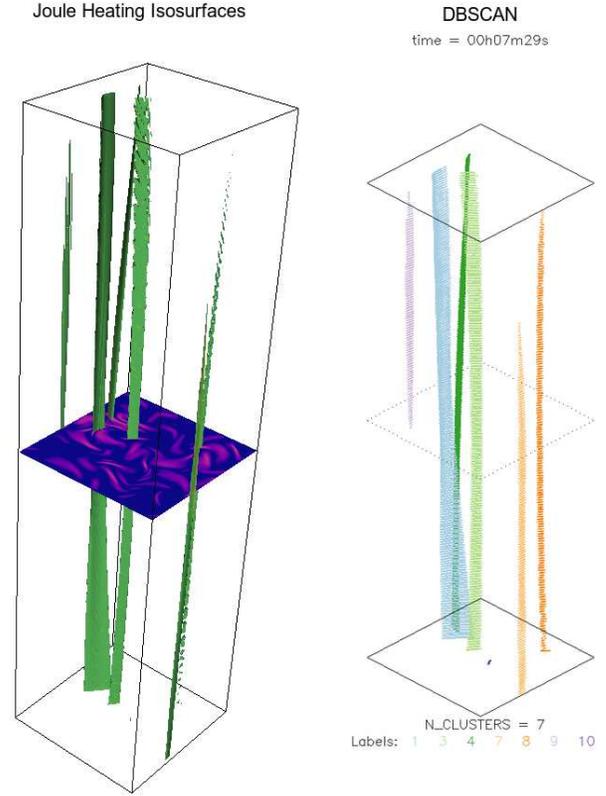}
\caption{\label{fig:highres}  Joule Heating isosurfaces ($3.0\times10^{-6}$) in
green with the corresponding XY slice cross-section and the DBSCAN events identifications
for one of the time steps in the high resolution HYPERION run.}
\end{figure}

Identifying the events in temperature space is relatively easy because the temperature
distribution changes rather smoothly within the box and the clustering algorithm
has no difficulty in identifying concentrations of points. The temperature changes,
however, are not a direct diagnostic of heating, but a consequence of it. Temperature
can change through direct heating, but also via thermal conduction from heating
deposited elsewhere.

\begin{figure}
\centering
\includegraphics[width=\linewidth]{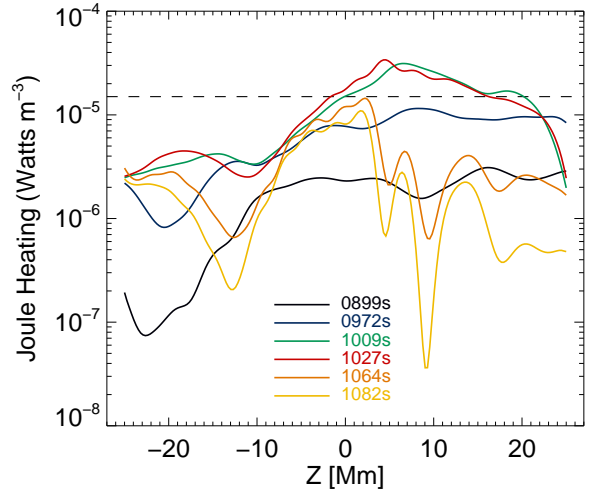}
\caption{\label{fig:jouleprof} Joule heating term evolution in time along the Z
direction at the location of the heating event highlighted with a blue circle in
Figure~\ref{fig:temperature}. The dashed line shows the threshold level used in
the event identification. Temperature profiles for that same location are shown
in Figure~\ref{fig:tempprof}.}
\end{figure}

For this reason, we look here at the differences
that arise when identifying events from the distribution of the Joule heating
term. This quantity is directly connected to the
energy released by the system and correlates with the Poynting flux changes
\citep{2015ApJ...submitted}.

\begin{figure*}
\centering
\includegraphics[width=\linewidth]{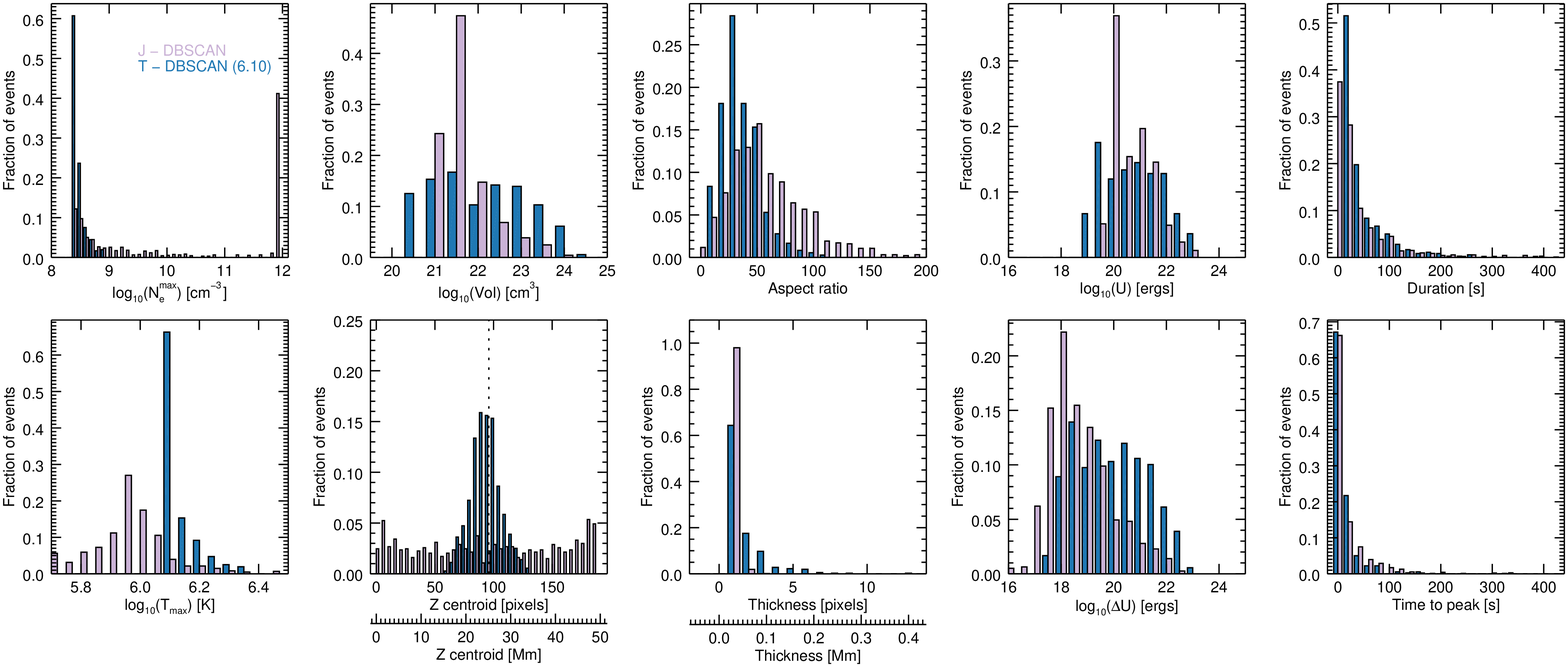}
\caption{\label{fig:comparison_to_joule} Properties of the Joule heating events identified
with DBSCAN as compared to the temperature heating events ($\tau_{logT}=6.10$) shown in
Figures~\ref{fig:comparison_methods} and \ref{fig:comparison_thresholds}. The number of
Joule heating events is 935.}
\end{figure*}

For the identification we choose a threshold value of $\tau_{Joule}=1.0\times 10^{-5} \,\rm Watts\,m^{-3}$
that corresponds to 2.7 times the standard deviation $\sigma$ of the electric current
distribution ($j_z=0.159 \rm A\,m^{-1}$ and $\eta=5.59\times10^{-5}$ ohm m), shown at
the top panel of Figure~\ref{fig:PDF} for one of the time steps. Figure~\ref{fig:events}
shows on the two right-most panels the Joule heating isosurfaces at that threshold
level and the DBSCAN events identification assuming $\epsilon=10$, $N_{pts}=10$.
The Joule heating term outlines the location of the current sheets where reconnection
takes place. The currents extend all along the box and
therefore the clustering algorithm identifies events that are longer than
the temperature events. Their configuration is also more complex. Being thinner,
longer and with peaks and troughs along their structure, results in a fractionation
of the long sheets that are sometimes identified as multiple events. The online
movie version of Figure~\ref{fig:events} shows the detections as a function of time.

This discretization of the structures is in part due to the noise introduced by the high Reynolds number respect to the
resolution at 128$\times$128x192. A
better resolved \citep[see][for a more complete discussion on this topic]{SerMat10, WanSer10}
run with twice the resolution across (256x256x192) shows smooth coherent structures running from footpoint to footpoint (Figure~\ref{fig:highres}).
The clusters are better sampled and easier to track by our identification algorithms. In this particular
case we looked at a threshold $\tau_{Joule}=3.0\times 10^{-6} \,\rm Watts\,m^{-3}$, also at the 2.7$\sigma$ level ($j_z=0.123 \rm A\,m^{-1}$ and $\eta=2.79\times10^{-5}$ ohm m).
The corresponding current distribution is shown at the bottom panel of Figure~\ref{fig:PDF}. Unfortunately, the duration of this run did not provide sufficient statistics
to be compared to the temperature or the Joule heating low resolution case. This will be subject of a future investigation. The observed trends reveal that the high resolution case exhibits thinner current sheets and higher currents peaks than the low resolution.
This point was also noted by \citet{baumann2013}, who similarly found current sheet widths depend on numerical resolution.
The remainder of this section discusses the results of the low resolution case.

\begin{figure}
\centering
\includegraphics[width=\linewidth]{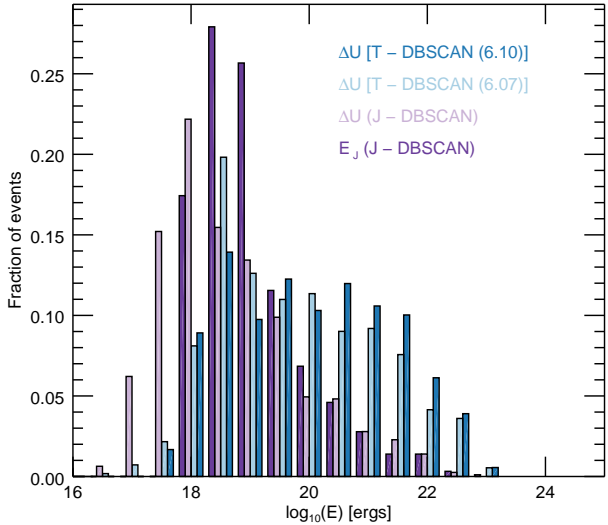}
\caption{\label{fig:joule_energy} Comparison of time integrated Joule Heating energy
and the internal energy increment for the events identified in the Joule Heating space (J).}
\end{figure}

Figure~\ref{fig:jouleprof} shows the Joule heating changes in time along the Z
direction at the same location as Figure~\ref{fig:tempprof} for the temperature.
There are similarities in that the temperature enhancement takes place towards the
top of the box (positive Z) which is also where the increase in Joule heating happens.
The peak in the Joule heating term takes place nearly simultaneously,
sometimes before in other events, to the peak in temperature,
which is something that we had already seen in HYPERION for the electric current
and maximum temperature \citep{2015ApJ...submitted}.

The properties for these events are shown in Figure~\ref{fig:comparison_to_joule}. The most striking differences have already been described. The Joule heating events
are longer and therefore their aspect ratio numbers are higher. The events are thinner, as expected for the size of the current sheets, although that is dependent
on the choice of threshold. The centroid positions along the Z direction show that in contrast to the temperature events, the Joule heating events are uniformly
distributed along the length of the box. The figure also shows that there are no significant differences in estimating the duration of the events when they are identified in temperature and Joule heating space. The maximum density in the Joule heating events shows a peak at $10^{12} \rm cm^{-3}$. This is not surprising because the events are extending to the footpoints where the density is higher. In the high resolution case where events are extended and do not break up in smaller fragments, we expect that distribution to skew almost completely to higher densities.

In terms of the energy distribution, our results indicate that there are differences
between the internal energies estimated for temperature events and those estimated for
Joule heating events. While both cover the same energy range, the distributions are
noticeably different. Section~\ref{sect:kolmogorov} discusses the shapes of the
distributions in more detail. The differences can be explained in terms
of volume. The internal energy is highly correlated with volume and the volume
distributions show similar discrepancies. The estimated volume depends on the
number of voxels in the cluster, which in turn depends on the thresholds. While
the temperature events seem to be a good proxy of where, across the flux tube the
heating is taking place, and for how long, the volume highlighted by the temperature
increment does not correspond to the volume where the heating takes place.

We find, however, that the internal energy increment ($\Delta U$) does not appear to
be a bad proxy for the heating of the events, provided we know where the heating is deposited.
Figure~\ref{fig:joule_energy} shows that for Joule events $\Delta U$, already shown in Figure~\ref{fig:comparison_to_joule},
compares well with the actual time integrated Joule Heating energy available in
the clusters ($E_j$). This distribution gives us the actual heating energies for the
events in this run, which are capable of producing temperatures up to 4 MK,
with the bulk under 2 MK. The distribution of $\Delta U$ for
the temperature events shows a flatter distribution, particularly for a higher
threshold in temperature (6.10), but reducing the threshold (6.07) and therefore reaching to
lower energy events, skews the distribution in the direction of those low energies
and brings it closer to the energy distribution in the Joule events.

\begin{table*}[t]
    \centering
    \begin{tabularx}{\textwidth}{c c c c c c c c }
    Case & $\mu$ & $\sigma$ & $D_{\text{max}}$ & $\alpha_{1\%}$ & $\alpha_{5\%}$ & $\alpha_{10\%}$ & KS Test Result  \\ \hline
    \rowcolor{cyan!20} Log-normal, DBSCAN, $\Delta U$ ($\log{T} > 6.07$) & $19.96 \pm 0.08$ & $1.33 \pm 0.05$ & 0.089 & 0.069 & 0.058 & 0.052 & Reject $H_{0}$ \\
    \rowcolor{cyan!80} Log-normal, DBSCAN, $\Delta U$ ($\log{T} > 6.10$) & $20.22 \pm 0.08$ & $1.33 \pm 0.06$ & 0.066 & 0.086 & 0.072 & 0.065 & Accept $H_{0}$ at 1\% and 5\% levels, reject 10\% \\
    \rowcolor{orange!20} Log-normal, PDC, $\Delta U$ ($\log{T} > 6.07$) & $20.49 \pm 0.14$ & $1.55 \pm 0.10$ & 0.075 & 0.101 & 0.084 & 0.076 & Accept $H_{0}$ \\
    \rowcolor{orange!80} Log-normal, PDC, $\Delta U$ ($\log{T} > 6.10$) & $20.69 \pm 0.14$ & $1.40 \pm 0.11$ & 0.071 & 0.114 & 0.095 & 0.085 & Accept $H_{0}$ \\
    \rowcolor{violet!20} Log-normal, DBSCAN, $\Delta U$ ($J$ events) & $18.89 \pm 0.06$ & $1.19 \pm 0.04$ & 0.081 & 0.058 & 0.048 & 0.043 & Reject $H_{0}$ \\
    \rowcolor{violet!70} Log-normal, DBSCAN, $E_{J}$ ($J$ events) & $19.29 \pm 0.04$ & $0.87 \pm 0.03$ & 0.143 & 0.053 & 0.044 & 0.040 & Reject $H_{0}$ \\

    \hline
    & & & & & & &  \\
    Case & $x_{\text{min}}$ & $b$ & $D_{\text{max}}$ & $\alpha_{1\%}$ & $\alpha_{5\%}$ & $\alpha_{10\%}$ & KS Test Result  \\ \hline
    \rowcolor{cyan!20} Power law, DBSCAN, $\Delta U$ ($\log{T} > 6.07$) & $2.03 \times 10^{18}$ & 1.24 & 0.427 & 0.072 & 0.060 & 0.054 & Reject $H_{0}$ \\
    \rowcolor{cyan!80} Power law, DBSCAN, $\Delta U$ ($\log{T} > 6.10$) & $1.47 \times 10^{21}$ & 1.59 & 0.109 & 0.162 & 0.135 & 0.122 & Accept $H_{0}$ \\
    \rowcolor{orange!20} Power law, PDC, $\Delta U$ ($\log{T} > 6.07$) & $8.64 \times 10^{20}$ & 1.42 & 0.181 & 0.162 & 0.132 & 0.121 & Accept $H_{0}$ at 1\% level, reject at 5 and 10\% \\
    \rowcolor{orange!80} Power law, PDC, $\Delta U$ ($\log{T} > 6.10$) & $2.27 \times 10^{20}$ & 1.35 & 0.196 & 0.150 & 0.125 & 0.113 & Reject $H_{0}$ \\
    \rowcolor{violet!20} Power law, DBSCAN, $\Delta U$ ($J$ events) & $8.95 \times 10^{18}$ & 1.40 & 0.154 & 0.090 & 0.075 & 0.068 & Reject $H_{0}$ \\
    \rowcolor{violet!70} Power law, DBSCAN, $E_{J}$ ($J$ events) & $8.81 \times 10^{18}$ & 1.50 & 0.066 & 0.071 & 0.059 & 0.053 & Accept $H_{0}$ at 1\% level, reject at 5 and 10\% \\

    \end{tabularx}
    \caption{MLE parameters of the empirical distributions of event energies and KS tests to determine consistency with log-normal distributions (top) or power law distributions (bottom).  \label{table:ks} }
    \parnotes
\end{table*}

\subsection{Kolmogorov-Smirnov Tests}
\label{sect:kolmogorov}
We discuss here the shape of the energy distributions in some detail.
The histograms visually appear consistent with log-normal distributions or perhaps power laws, so to determine their consistency with these distributions we have performed Kolmogorov-Smirnov (KS) tests.  KS tests work by comparing the cumulative distribution functions (CDFs) of the empirical distribution to a theoretical distribution, and measuring their maximum separation $D_{\text{max}}$.  If $D_{\text{max}}$ is greater than a critical value, then we reject the null hypothesis and conclude that the empirical distribution is not consistent with the theoretical one.

To begin, we first perform maximum likelihood estimation (MLE) of the parameters of the distributions: mean $\mu$ and standard deviation $\sigma$ for a log-normal distribution, and the minimum value $x_{\text{min}}$ and slope $b$ for a power law.  In the former case, the MLE of the mean and standard deviation can be calculated directly.  In the latter case, the slope $b$ can be calculated directly for some assumed $x_{\text{min}}$, but the $x_{\text{min}}$ value itself needs to calculated through other means.  To calculate $x_{\text{min}}$, we follow the method described by \citet{clauset2007}, which is a KS minimization technique.  For each value of $x_{i}$ in the data set, assume that $x_{\text{min}} = x_{i}$, calculate the MLE of the slope, then perform a KS test on that fitted power law for all $x_{i} \ge x_{\text{min}}$.  The power law with the smallest $D_{\text{max}}$ gives the best estimate of $x_{\text{min}}$.  This method does not provide uncertainties for the power law parameters, however.

The MLE parameters and KS test results are shown in Table \ref{table:ks}.  The result depends on the variable used for clustering, the threshold, and the clustering method.  When clustering based on temperature, DBSCAN tends to find distributions consistent with power laws, while PDC finds distributions consistent with log-normal.  When clustering based on the Joule heating, the distributions appear to be neither consistent with power laws nor log-normal.  Visually, these distributions do appear to have a noticeable skew to them.  Because the results depend intimately on the choices in the clustering methods, we emphasize again that the derived values of heating are only estimates of the true energy release.

\section{Discussion}

In this paper we report the results of a detailed analysis of the data obtained running the HYPERION code on a specific case which represents a model of a typical coronal loop. We have assumed a loop base density $n_{base}=10^{17}\,$ m$^{-3}$ (producing a loop apex density $n_{apex}\approx3\times10^{14}\,$ m$^{-3}$) , a loop base temperature $T_{base}= 2\times10^{4}\,$  K, perpendicular dimension $L_{\perp} = 4 \times 10^{6}\,$  m and a loop length of $L_0$= 50,000 km.   A basic loop magnetic field strength of $B=0.01$ Teslas is used.
The dimensionless time scale of the forcing, $t^*$, is set to a value that corresponds to a 5 minute convection time scale and the normalized
driving velocity $V_{phot}$ is $10^3$~m~s$^{-1}$.  Simulations were performed at two numerical resolutions.  For the first simulation, the resolution is $128\times128\times192$  (the 128 refers to resolution in the planes,  and the 192 refers to the central difference grid pointsin the axial direction).  At this resolution we were able  to run the code for almost one hour of real time, a time sufficient to produce a large enough number of events to allow a statistical analysis.  At this resolution the physical grid size is  31 km in the perpendicular direction and 260 km in the parallel one.  We also performed a second simulation, at lower physical dissipation, with a resolution of $256\times256\times192$.  This  simulation was run for a physical time of about fifteen minutes, which was long enough to verify that the results show the same sort of trends as have been seen in the RMHD simulation as the resolution is increased.

At the beginning of the simulation the loop is in hydrostatic equilibrium, as specified in Section \ref{sect:model}, and the thermodynamical equilibrium is ensured by ramping up conduction and radiation from zero, so that the loop temperature does not collapse before the reconnection-driven heating mechanism becomes efficient within the elementary events. Note that there is no ``ad hoc'' heating function -- the only heating is due to  the dissipation of the currents induced by the photospheric motions. The resulting loop is a multi-temperature system with peaks over 4MK within the elementary events and and lows well under 1 MK elsewhere.

Adopting the machine learning algorithms described in the previous sections, we have been able to identify elementary events occurring continuously during the evolution of the system driven by the boundary motions. We have measured their spatial distribution and other important properties, namely:

1) the size and duration of the events.

2) the energy released, evaluated both as the variation of internal energy and also as the energy due to current dissipation within the events.

3) the variation of density with respect to the initial conditions.

The results detailed in this paper confirm the fundamental role of elementary events in the process of  dissipating energy transferred from the photosphere into the corona. The simulations also demonstrate that the release of energy in the corona occurs through the dissipation of currents localized in current sheets resulting from the nonlinear coronal dynamics triggered by the large scale photospheric forcing. Their existence appears clearly in Figures \ref{fig:events}  and \ref{fig:highres} which show (at both resolutions) that they extend along the entire loop and occupy a volume very small compared with the volume of the loop. From Figures \ref{fig:comparison_methods}, \ref{fig:comparison_thresholds} and \ref{fig:comparison_to_joule} it appears that their thickness is of a few pixels, which in dimensional units means less than 100 km. The thickness extending for a few pixels remains true also at higher resolution, less than 50 km. This fact can be easily understood considering that the current sheets are the manifestation of the energy cascade toward small spatial scales induced by the nonlinear effects dominating the perpendicular dynamics
\citep{2008ApJ...677.1348R}. Typical of turbulent cascades, the spectral energy flow from large to small scales is stopped at the dissipative scale where dissipation dominates and provides a sink for the energy flow: here the dissipation rate equals the nonlinear transfer rate leading to a dissipative scale (and therefore current sheets thickness) decreasing as a power of the Reynolds number \citep[ e.g.,][]{biskamp03}.
The conclusion is that the thickness of the current sheets decreases with increasing resolution and in the solar conditions is far smaller than any possible observational resolution.

The current sheets form and disrupt, generating a localized increase of the plasma temperature (the elementary event), on time scales determined by the nonlinear dynamics induced by the growth of the component of the magnetic field ($b_{\perp}$) perpendicular to the main field. Figures \ref{fig:comparison_methods}, \ref{fig:comparison_thresholds} and \ref{fig:comparison_to_joule} show that the duration of each event is very rarely longer than 100 seconds, up to 200 seconds, and in most cases varies between a few and 50 seconds.

In the simulations the energy dissipated in the current sheets is transformed into internal energy through Joule heating. In this paper we measure the amount of energy involved in each event in two ways: (1) by determining the variation of internal energy during the occurrence of the event and (2) by computing the Joule heating, $\eta j^2 $, created within the event by the current dissipation. We have identified the currents within the current sheets as those which form the non-Maxwellian tail.  This is clearly seen in Figure \ref{fig:PDF}, which shows the probability distribution function (PDF) of electric currents calculated at one time during the evolution. The variation of internal energy method and the Joule heating method are in good agreement in the evaluation of the energies associated with the elementary events, exhibiting  a distribution centered around $10^{18}$ - $10^{21}$ ergs,  (see Figures \ref{fig:comparison_methods}, \ref{fig:comparison_thresholds} and \ref{fig:comparison_to_joule}). Few events are found at higher energies up to $10^{23}$ ergs.

Looking at the PDFs of the electric current density, it is interesting to note that both the amount of currents in the tail of the distribution and their values increase considerably with resolution. At the higher resolution shown in this paper the currents are more concentrated within the currents sheets and also the number of currents sheets increases and their thickness decreases. We expect this trend to continue when resolution, and the corresponding Reynolds number, are further increased, see \cite{Rappazzo2017}, keeping the total energy associated with the events in the same range $10^{18}$ - $10^{21}$ ergs. This point must be verified by running much higher resolution simulations.

We also have found a good correlation between the energies, durations and volumes of the events. The few events in the high energy tail of the distribution are also in the long duration range and big volume range.

As mentioned in Section 5.2, Joule heating events extend all the way to the bottom of the loop where the density is the typical density above the chromosphere. They are much longer than the temperature events because of the density stratification,
nevertheless the much higher density at the bottom does not allow to raise the temperature as significantly as at the loop apex with a similar local current dissipation rate.
They also are thinner than the temperature events, a property related to the magnetic connectivity associated with current sheets.
Recently \cite{Rappazzo2017} have traced at a given time magnetic field
lines from current sheets in a high resolution reduced MHD simulation of a coronal loop
with geometry and parameters similar to those shown in this paper. They have
found that the field lines traced from such current sheets (elongated in the
axial direction) fill a larger volume around the current sheets. As thermal conductivity is highly anisotropic and heat transport occurs essentially along magnetic field lines, it is clear that at each instant of time heating from current sheets is transported in
these regions around current sheets. These are therefore the regions where heating occurs and our temperature events are found.

The density profile of the loop is practically unaltered (up to 5\%) during the evolution. The fact that the maximum density within the temperature events has a distribution is mainly related to the fact that they extend down from the top of the loop where the density increases due to the stratification. Since the Joule heating events extend all the way to the bottom of the loop explains the appearance of a peak in the maximum density in their analysis.
The small amount of energy involved in each event and the fact that the current sheets move across the loop during the current dissipation along with the mentioned spread of conductive flux has as a consequence that very little thermal energy is conducting from the apex to the bottom of the loop. It follows that the evaporation from the chromosphere does not affect significantly the density of the loop.

As the numerical resolution is increased (and the physical dissipation decreased) we expect to find higher currents within the current sheets with smaller thickness and therefore smaller volumes. This is confirmed by the comparison of low vs. high resolution simulations presented in this paper. We also expect an increase of the number of elementary events occurring during the same time. The energies involved should remain more or less the same because of the simultaneous increase of the current densities and decrease of the volumes. We will analyze Joule heating events in high-resolution RMHD
simulations and compare them with the statistics presented here in an upcoming work.

It is important to emphasize that in our MHD framework all the dissipated energy goes into heating and the particle acceleration due to the well known Fermi mechanisms (stochastic and systematic) are neglected. These mechanisms are at work at the very small scales the present paper indicates exist in the solar corona, contributing to the formation of high energy tail in the particle distribution
(see \cite{vlahos} and references therein)

The results of this paper indicate that it will still remain very challenging to observationally resolve elementary events because of their typical size and duration and the small amount of energy involved in each of them. At present only indirect observational tools will be able to verify the theory, as we have done in a previous paper through, for example, the use of emission measures. Moreover, a correct theoretical approach requires the use of 3D equations in order properly to take into account the perpendicular dynamics which are responsible for both the energy deposition and the continuous change of magnetic connectivity which makes the energy exchange between corona and chromosphere possible.

\acknowledgments

This material is based upon work supported by the National Aeronautics and Space Administration under Grant/Contract/Agreement No.
NNH17AE96I issued through the Heliophysics Grand Challenge Research Program. RBD was also supported by the Naval Research
Laboratory 6.1 program. F. Rappazzo and M. Velli acknowledge support of the  NASA DRIVE SOLFer grant N. 80NSSC20K0627. 
Computer simulations were performed on the LCP\&FD Intel Core i7 cluster and the LCP\&FD AMD Magny Cours cluster. We thank H.P.
Warren for many helpful discussions, and the referee for his useful remarks.


\begin{thebibliography}{}
\expandafter\ifx\csname natexlab\endcsname\relax\def\natexlab#1{#1}\fi


\bibitem[Aschwanden(2010)]{aschwanden2010}Aschwanden, M.J.: 2010, {\it Solar Physics} {\bf 262}, 235.

\bibitem[Aschwanden \& Peter(2017)]{2017ApJ...840....4A} Aschwanden, M.~J., \& Peter, H.\ 2017, \apj, 840, 4

\bibitem[Baumann et al.(2013)]{baumann2013}
Baumann, G., {Galsgaard}, K., \& {Nordlund}, {\AA} 2013, Solar Phys., 284, 467

\bibitem[Berger (1991)]{Berger 1991}
Berger,M. A. 1991, A\&A, 252, 369

\bibitem[Bingert \& Peter (2011)]{Bingert}
Bingert, S., \& Peter, H. 2011, A\&A, 530, A112

\bibitem[{{Biskamp}(2003)}]{biskamp03}
{Biskamp}, D. 2003, {Magnetohydrodynamic Turbulence} (Cambridge: Cambridge
University Press)

\bibitem[{Biskamp \& Welter (1989)}]{Biskamp & Welter 1989}
Biskamp, D., \& Welter, H. 1989, Physics of Fluids, B1, 1964

\bibitem[\protect\citeauthoryear{Bobra \& Mason}{2018}]{bobra2018} Bobra M.~G., Mason J.~P., 2018, mlsd.book


\bibitem[Brooks et al.(2012)]{2012ApJ...755L..33B} Brooks, D.~H., Warren, H.~P., \& Ugarte-Urra, I.\ 2012, \apjl, 755, L33

\bibitem[Brooks et al.(2013)]{2013ApJ...772L..19B} Brooks, D.~H., Warren, H.~P., Ugarte-Urra, I., \& Winebarger, A.~R.\ 2013, \apjl, 772, L19

\bibitem[{Carpenter \& Kennedy(1994)}]{carpenter1994fourth}
Carpenter, M.~H., \& Kennedy, C.~A. 1994, Technical Report 109112

\bibitem[Caballero and Aranda(2013)]{caballero2013}Caballero, C., and Aranda, M.C.: 2013, {\it Solar Physics} {\bf 283}, 691.

\bibitem[Chiuderi(1993)]{Chiuderi 1993} Chiuderi, C. 1993, in ``Scientific Requirements for Future  Solar-Physics Space
Missions,'' 1993, eds. P. Maltby  and  B. Battrick, (ESA SP-1157),  p.25.

\bibitem[Clauset et al.(2007)]{clauset2007}
Clauset, A., Rohilla Shalizi, C., \& Newman, M.E.J.\ 2007 SIAM Rev., 51(4), 661-703.

\bibitem[{{Culhane} {et~al.}(2007){Culhane}, {Harra}, {James}, {Al-Janabi},
  {Bradley}, {Chaudry}, {Rees}, {Tandy}, {Thomas}, {Whillock}, {Winter},
  {Doschek}, {Korendyke}, {Brown}, {Myers}, {Mariska}, {Seely}, {Lang}, {Kent},
  {Shaughnessy}, {Young}, {Simnett}, {Castelli}, {Mahmoud}, {Mapson-Menard},
  {Probyn}, {Thomas}, {Davila}, {Dere}, {Windt}, {Shea}, {Hagood}, {Moye},
  {Hara}, {Watanabe}, {Matsuzaki}, {Kosugi}, {Hansteen}, \&
  {Wikstol}}]{2007SoPh..243...19C}
{Culhane}, J.~L., {Harra}, L.~K., {James}, A.~M., {et~al.} 2007, \solphys, 243,
  19

\bibitem[Dahlburg et al. (2016b)]{Dahlburg2016b}
Dahlburg, R. B., Laming, J. M., Taylor, B. D., \& Obenschain, K.
2016b, \apj, 831, 160

\bibitem[{{Dahlburg} {et~al.}(2012){Dahlburg}, {Einaudi}, {Rappazzo}, \&
  {Velli}}]{Dahlburg 2012}
{Dahlburg}, R.~B., {Einaudi}, G., {Rappazzo}, A.~F., \& {Velli}, M. 2012, \aap,
  544, L20

\bibitem[{Dahlburg} {et~al.}(2016)]{2015ApJ...submitted}
{Dahlburg}, R.~B., {Einaudi}, G., {Taylor}, B., {Ugarte-Urra}, I., {Warren}, H. P.,
{Rappazzo}, A.~F., \& {Velli}, M. 2016, \apj, 817, 47, Paper I

\bibitem[{Dahlburg} {et~al.}(2018)]{2018ApJ...868.116D}
{Dahlburg}, R.~B., {Einaudi}, G., {Ugarte-Urra}, I., {Rappazzo}, A.~F., \& {Velli}, M. 2018, \apj, 868, 116

\bibitem[{{Dahlburg} \& {Picone}(1989)}]{1989PhFlB...1.2153D}
{Dahlburg}, R.~B., \& {Picone}, J.~M. 1989, Physics of Fluids B, 1, 2153

\bibitem[{{Dahlburg} {et~al.}(1986){Dahlburg}, {Zang}, \&
  {Montgomery}}]{1986JFM...169...71D}
{Dahlburg}, R.~B., {Zang}, T.~A., \& {Montgomery}, D. 1986, Journal of Fluid
  Mechanics, 169, 71

  \bibitem[{{Dere} {et~al.}(1997){Dere}, {Landi}, {Mason}, {Monsignori Fossi}, \&
  {Young}}]{1997AAS...125...149}
{Dere}, K.~P., {Landi}, E., {Mason}, H.~S., {Monsignori Fossi}, B.~C., \& {Young}, P.~R.
  1997, Astron. Astrophys. Suppl. Ser., 125, 149.

 \bibitem[{Diamond \& Biskamp (1990)}]{Diamond & Biskamp 1990}
 Diamond, P. H., \& Biskamp, D. 1990, Physics of Fluids, B2, 681

 \bibitem[{Dmitruk} {et~al.} (1998)]{Dmitruk 1998}
 Dmitruk, P. ,Gomez D. O., and DeLuca,E. E. 1998 \apj, 505, 974

\bibitem[{{Dorch} \& {Nordlund}(2001)}]{2001A&A...365..562D}
{Dorch}, S.~B.~F., \& {Nordlund}, {\AA}. 2001, \aap, 365, 562

\bibitem[{Einaudi \& Velli(1994)}]{Einaudi & Velli 1994}
Einaudi, G. \& Velli, M., 1994, in "Advances in Solar Physics``, ed. G. Belvedere,
M. Rodono`, \& G. M. Simnett (Berlin: Springer-Verlag), 149

\bibitem[Einaudi et al.(1996)]{1996ApJ...457L.113E}
Einaudi, G., Velli, M., Politano, H., \& Pouquet, A. 1996, \apjl, 457, L113

\bibitem[Ester et al.(1996)]{ester1996}{Ester}, M., {Kriegel},H.-P., {Sander}, J., \& {Xu}, X. 1996, AAAI, pages 226--231.

\bibitem[Furth et al.(1963)]{Furth}
Furth, H. P., Killeen, J., \& Rosenbluth, M. N. 1963, Physics of Fluids, 6, 459

\bibitem[{{Galsgaard} \& {Nordlund}(1996)}]{1996JGR...10113445G}
{Galsgaard}, K., \& {Nordlund}, {\AA}. 1996, \jgr, 101, 13445

\bibitem[Georgoulis et~al.(1998)]{Georgoulis}
{Georgoulis}, M.~K., {Velli}, M., \& {Einaudi}, G. 1998, \apj, 497, 957

\bibitem[{Gold \& Hoyle (1960)}]{Gold & Hoyle 1960}
Gold, T., \& Hoyle, F. 1960, MNRAS, 120, 89

\bibitem[Gomez \& Ferro Fontan (1988)]{GFF}
Gomez, D. and \& Ferro Fontan, C. 1988, Sol. Ph., 116, 33

\bibitem[{{Gudiksen} \& {Nordlund}(2002)}]{2002ApJ...572L.113G}
{Gudiksen}, B.~V., \& {Nordlund}, {\AA}. 2002, \apjl, 572, L113

\bibitem[Guerreiro et al.(2017)]{GuHan17}
Guerreiro, N., Haberreiter, M., Hansteen, V. H., \& Schmutz, W. 2015, \aap, 603, A103

\bibitem[Hansteen et al.(2010)]{HanHa10}
Hansteen, V. H., Hara, H. , De Pontieu, B., \& Carlsson, M. 2010, \apj, 718, 1070

\bibitem[Hansteen et al.(2015)]{HanGu15}
Hansteen, V. H., Guerreiro, N., De Pontieu, B., \& Carlsson, M. 2015, \apj, 811, 106

\bibitem[Hendrix \& Van Hoven (1996)]{Hendrix & Van Hoven 1996}
Hendrix, D. L., \& Van Hoven, G. 1996, ApJ, 467, 887

\bibitem[Landi et al.(2012)]{2012ApJ...744...99L}
Landi, E., Del Zanna, G., Young, P.~R., et al.\ 2012, \apj, 744, 99. doi:10.1088/0004-637X/744/2/99

\bibitem[Longcope \& Sudan (1994)]{Longcope & Sudan 1994}
Longcope, L., \& Sudan, L. 1994, ApJ, 437, 491

\bibitem[Martens et al.(2012)]{martens2012}Martens, P.C.H., Attrill, G.D.R., Davey, A.R., Engell, A., Farid, S., Grigis, P.C., Kasper, J., Korreck, K., Saar, S.H., Savcheva, A., Su, Y., Testa, P., Wills-Davey, M., and, ...: 2012, {\it Solar Physics} {\bf 275}, 79.

\bibitem[{{Meyer} {et~al.}(2012){Meyer}, {Balsara}, \&
  {Aslam}}]{2012MNRAS.422.2102M}
{Meyer}, C.~D., {Balsara}, D.~S., \& {Aslam}, T.~D. 2012, \mnras, 422, 2102

\bibitem[{{Mikic} {et~al.} (1989)}]{Mikic 1989}
Mikic, Z., Schnack, D. D., \& Van Hoven, G. 1989, ApJ, 338, 1148

\bibitem[{{Parker}(1972)}]{1972ApJ...174..499P}
{Parker}, E.~N. 1972, \apj, 174, 499

\bibitem[{{Parker}(1983)}]{Parker 1983}
{Parker}, E.~N. 1983, ApJ, 264, 642

\bibitem[{{Parker}(1988)}]{1988ApJ...330..474P}
{Parker}, E.~N. 1988, \apj, 330, 474

\bibitem[Parker (1989)]{Parker 1989}
{Parker}, E.~N. 1989, Sol. Phys., 121, 271

\bibitem[Parker (1991)]{Parker 1991}
{Parker}, E.~N. 1991, ApJ, 372, 719

\bibitem[{{Parker}(1994)}]{1994ISAA....1.....P}
{Parker}, E.~N. 1994, {Spontaneous current sheets in magnetic fields} (New York: Oxford
  University Press)

\bibitem[Pedregosa et al.(2011)]{pedregosa2011} Pedregosa, F., Varoquaux, G., Gramfort, A., et al., 2011, Journal of Machine Learning Research, 12, 2825

\bibitem[Peter et al.(2013)]{peter2013} Peter, H., Bingert, S., Klimchuk, J.~A., et al.\ 2013, \aap, 556, A104

\bibitem[Rappazzo(2015)]{2015ApJ...815....8R}
Rappazzo, A. F. 2015, \apj, 815, 8

\bibitem[Rappazzo et al.(2018)]{2018MNRAS...submitted}
Rappazzo, A. F., Dahlburg, R. B.,  Einaudi, G., \& Velli, M., 2018, \mnras, 478, 2257

\bibitem[Rappazzo et al. (2017)]{Rappazzo2017}
Rappazzo A.F., Matthaeus, W.H., Ruffolo, D., Velli, M., \& Servidio, S.
2017, \apj, 844, 87

\bibitem[Rappazzo et al.(2007)]{2007ApJ...657L..47R}
Rappazzo, A. F., Velli, M., Einaudi, G., \& Dahlburg, R. B., 2007, \apjl, 657, L47

\bibitem[Rappazzo et al.(2008)]{2008ApJ...677.1348R}
Rappazzo, A. F., Velli, M., Einaudi, G., \& Dahlburg, R. B. 2008, \apj, 677, 1348

\bibitem[Rappazzo et al.(2010)]{2010ApJ...722...65R}
Rappazzo, A. F., Velli, M., \& Einaudi, G. 2010, \apj, 722, 65

\bibitem[Rappazzo \& Velli(2011)]{2011PhRvE..83f5401R}
Rappazzo, A. F., \& Velli, M. 2011, Phys. Rev. E, 83, 065401

\bibitem[Rappazzo et al.(2013)]{2013ApJ...771...76R}
Rappazzo, A. F., Velli, M., \& Einaudi, G. 2013, \apj, 771, 76

\bibitem[Rappazzo \& Parker(2013)]{2013ApJ...773L...2R}
Rappazzo, A. F., \& Parker, E. N. 2013, \apj, 773, L2

\bibitem[Rempel(2017)]{rem17}
Rempel, M. 2017, \apj, 834, 10

\bibitem[{{Reale}(2014)}]{2014LRSP...11....4R}
{Reale}, F. 2014, Living Reviews in Solar Physics, 11, 4

\bibitem[Schmelz et al.(2014)]{2014ApJ...795..171S}
Schmelz, J.~T., Pathak, S., Brooks, D.~H., Christian, G.~M., \& Dhaliwal, R.~S.\ 2014, \apj, 795, 171

\bibitem[Servidio et al.(2010)]{SerMat10} Servidio, S., Matthaeus, W. H., Shay, M. A., Dmitruk, P., Cassak, P. A. \& Wan, M. 2010, Phys.\ Plasmas 17, 032315

\bibitem[Strauss (1976)]{Strauss 1976}
Strauss, H. 1976, Physics of Fluids, 19, 134

\bibitem[Strauss (1993)]{Strauss 1993}
Strauss, H. 1993, Geophys. Res. Lett., 20, 325

\bibitem[Sturrock \& Uchida (1981)]{Sturrock & Uchida 1981}
Sturrock, P. A., \& Uchida, Y. 1981, ApJ, 246, 331

\bibitem[Van Ballegooijen (1986)]{Van Ballegooijen 1986}
Van Ballegooijen, A. A. 1986, ApJ, 311, 1001

\bibitem[Vlahos \& Isliker (2019)]{vlahos}
Vlahos, L., \& Isliker, H. 2019, Plasma Physics and Controlled Fusion, 61,1, (014020)

\bibitem[Wan et al.(2010)]{WanSer10} Wan, M., Oughton, S., Servidio, S., \& Matthaeus, W. H. 2010, Phys.\ Plasmas 17, 082308

\bibitem[{{Warren} {et~al.}(2008){Warren}, {Ugarte-Urra}, {Doschek},
{Brooks} \& {Williams}}]{2008ApJ...686..131L}
{Warren}, H., {Ugarte-Urra}, I., {Doschek}, G.~A., {Brooks}, D.~H.,
\& {Williams}, D.~R. 2008, \apj, 686, L131

\bibitem[Xie et al.(2017)]{2017ApJ...842...38X} Xie, H., Madjarska, M.~S., Li, B., et al.\ 2017, \apj, 842, 38

\end{thebibliography}
\end{document}